\documentclass[onecolumn]{article}    % Enable this line and disable the 
\usepackage{graphicx}          % Include this line if your 
\usepackage{amssymb}
\usepackage{amsthm}
\usepackage{subfig}
\usepackage{mathtools}
\usepackage{authblk}
\usepackage{hyperref}

\usepackage{bm}
\usepackage{amsmath}
\usepackage{amsfonts}
\usepackage{subfig}

\usepackage{mathabx}
\usepackage{blindtext}

\usepackage{multirow}

\usepackage[a4paper, rmargin = 2.5cm, lmargin = 2.5cm]{geometry}

\begin{document}

\title{Study on Autonomous Gravity-assists with a Path-following Control }

\author{Rodolfo B. Negri\thanks{PhD Candidate, Division of Graduate Studies, National Institute for Space Research - INPE, São José dos Campos, Brazil.},
\ and Antônio F. B. de A. Prado\thanks{Pro-Rector of the Graduate School, National Institute for Space Research - INPE, São José dos Campos, Brazil.}
}

\maketitle{}

\begin{abstract}
We investigate the autonomous control of gravity-assist hyperbolic trajectories using a path following control law based on sliding mode control theory. This control strategy ensures robustness to bounded disturbances. Monte Carlo simulations in the environments of Titan and Enceladus, considering significant insertion errors on the order of 50 km, demonstrate the effectiveness of the proposed approach. The Enceladus example showcases the applicability of the control strategy for close flybys of asteroids and small moons during scientific observations. It successfully stabilizes the orbital geometry within a short time span, avoiding collisions and enabling a close approach to Enceladus' surface with a separation distance of 10 km. Furthermore, we explore its application in a Jovian tour, considering a more complex N-body problem. Results indicate that the control system, while unable to guarantee a complete tour, plays a crucial role in ensuring precise trajectory control during flybys. In such cases, the vehicle guidance system requires higher precision than what can be achieved with a patched conics model. These findings demonstrate the effectiveness of the proposed control strategy for gravity-assist maneuvers and highlight its potential for various space exploration missions involving close encounters with celestial bodies.
\end{abstract}

\section{Introduction}

The interest in space exploration has exceeded scientific aims and increasingly reached economical domains. The expectancy is that this trend continues to grow, which would, with a high probability, imply in a tremendous amount of different space missions designed to explore every part of the solar system. This demand poses a great challenge for the ground operation facilities on Earth. A convenient solution is to increase the autonomous capabilities of the current spacecraft, eliminating or reducing the necessity of the ground in the loop. In addition, autonomous operations can imply in other advantages, such as reducing operational costs and time. 

Recent studies have focused in the automation of different aspects of space missions, but, as far as the authors know, no work has studied an important aspect of many interplanetary missions, which is the gravity-assist. Current gravity-assists operations consist in the ground team accessing the spacecraft state and uploading corrective maneuvers before and after the passage to guarantee the designed flyby trajectory. In many cases, this process does not imply in concerning limitations for the mission. However, as pointed out by Reference~\cite{quadrelli2015guidance}, outer planets tours are greatly affected by this operational profile and an autonomous operation would result in great benefit, as in a: 1) rapid turnaround and post-flyby cleanup; 2) successive and safer low altitude gravity-assists; 3) more efficient outer planet orbit insertion; 4) increase in the number and frequency of gravity-assists; 5) less propellant mass required. 

In this sense, this works intends to assess an important aspect of this autonomous operation, which is the control law. The control law for such operation is very likely to be a path following than a reference tracking law. A path following control is concerned in driving the vehicle to a trajectory geometry, with no time parameterization for the movement on the path. On the other hand, a trajectory tracking would make the vehicle converge to a specific point of the trajectory for a predefined time. In the gravity-assist context, the time is only an important variable in the transfer to the body where the swing-by will be performed. Once the encounter with the body takes place, the important aspect is to preserve the desired hyperbole, with no need to spend fuel in avoiding a little delay or advance in an hypothetical predefined time.

We apply a robust Keplerian path following law recently derived~\cite{negri2020novel} that showed promising results for small body missions~\cite{negri2021autonomous}. This path following lies on sliding mode control theory to guarantee robustness to bounded disturbances, which in multiple body systems, such as the outer planetary systems, could be important. Robustness would also be a desirable feature considering low-altitude flybys in environments such as the ones in Enceladus~\cite{lorenz2018enceladus}, Io~\cite{lorenz2015io} and Titan~\cite{pelletier2006atmospheric}. Although the drag levels found in these cases are generally low, robustness can guarantee a safer lower altitude for the flyby and allow a less cautious approach, at least under the trajectory perspective, when facing uncertainties in the environment. We run Monte Carlo simulations for Titan and Enceladus flybys considering a reasonable large encounter error. Finally, we consider a Jovian tour in order to analyze the control in the context it is most likely to be applied, outer planetary systems tours.

\section{Dynamics}

The gravity-assist concept is very simple. Its goal is to change the trajectory of a spacecraft about a main body through the encounter with an intermediate body that also orbits the main one (e.g. approaching a planet's moon to change the orbit about the planet). In this way, our study can be separated into two parts. One considering the hyperbolic trajectory relative to the body where the gravity-assist takes place, and others concerned with the general context where the autonomous gravity-assist will be applied, in a multiple body dynamical environment. The first part of this section describes the models applied to the first case, while the remainder is concerned in presenting the alternatives for obtaining and describing a trajectory in an N-body system.

\subsection{Hyperbolic Trajectory}
\label{sec:gen_dyn}

The equations of motion of the spacecraft relative to the body where the gravity-assist is performed can be written as:
\begin{subequations}
\begin{align}
\dot{\vec{r}} &= \vec{v}, \\
\dot{\vec{v}} &= \vec{f} + \vec{d} + \vec{u} ,
\end{align}
\end{subequations}
in which $\vec{r}$ represents the position relative to the body, and $\vec{v}$ the velocity. The function $\vec{f}$ represents the known dynamics of the system, unknown bounded disturbances are represented by $\vec{d}$, while the control command is written as $\vec{u} $.

In order to stress the control law, we assume that the only known dynamics are the point mass gravitational acceleration $\vec{f} = - \frac{\mu}{r^3} \vec{r}$, in which $\mu$ is the gravitational parameter of the gravity-assist body. The third-body effects of the main body are assumed as disturbances, taking the following form:
\begin{equation}
\label{eq:d3B}
\vec{d}_{3B} = - \mu_\Sun \left( \frac{\vec{R}}{R^3}  - \frac{\vec{R}_\Earth}{R_\Earth^3}  \right),
\end{equation}
where $\mu_\Sun$ stands for the gravitational parameter of the main body, $\vec{R}$ denotes the position of the spacecraft relative to the main body, and $\vec{R}_\Earth$ the position of the gravity-assist body relative to the main body.

In a close approach to Titan, the spacecraft experiences drag from Titan's atmosphere. Although this drag is of little impact for hyperbolic trajectories as close as the ones of Cassini (minimum of 880 km altitude)~\cite{pelletier2006atmospheric}, it can be more significant for closer approaches. Drag can also be experienced in a pass through Io and Enceladus' plumes~\cite{lorenz2015io,lorenz2018enceladus}, yet of little impact in a hyperbolic trajectory also.  Nevertheless, we will consider as a second source of disturbance the drag acceleration:
\begin{equation}
\label{eq:dD}
\vec{d}_{D} = - \frac{1}{2 m} C_D \rho A v \vec{v},
\end{equation}
for $C_D=2.2$ representing the drag coefficient, $m$ is the spacecraft's mass, which we will assume as 1,000 kg throughout this work, $\rho$ is the mass density, and $A$ is the cross-sectional area projected in the $\vec{v}$ direction, assumed here as $A=18.6$ m$^2$.

\subsection{Multiple body Dynamics}
\label{sec:multi}

Since the first works of Tsander proposing gravity-assists~\cite{tsander1964problems}, and the pioneering of Crocco~\cite{crocco1956one} in what can be considered a tour design~\cite{negri2020historical}, the gravity-assist concept has largely evolved, with applications in most of the interplanetary missions, and complex tours as the ones designed for the Europa clipper mission in the Jovian system~\cite{campagnola2019tour}. In this section, we will describe the models to obtain and simulate the trajectory of the spacecraft in the condition where an autonomous control for a gravity-assist is most valuable, that is to the application in outer planetary systems tours~\cite{quadrelli2015guidance}.

\subsubsection{Zero-SOI Patched Conics Tour Design}
\label{sec:zeroSOI}

The first multiple body model presented here is the simplest one, the Zero-SOI Patched Conics (0SOI-PC).  In this approach, it is assumed that the point where the spacecraft meets the gravity-assist body, in its trajectory about the main body, is exactly the position of the gravity-assist body in the main body reference frame. That is the reason of the ``zero-SOI'' nomenclature, as the magnitude of the Sphere of Influence (SOI) of the gravity-assist body is assumed to be small enough to hold the approximation. The zero-SOI approximation allows constraining much of the variables of the problem. This is specially useful for optimization routines, as the decision variables are greatly reduced, which is our case. 

We assume that the time $t_j$ for each $j$-th gravity-assist is a decision variable, as well as the periapsis $r_{pj}$ of the hyperbolic trajectory about the gravity-assist body. The position and velocity of the gravity-assist body, respectively $\vec{R}_{\Earth_j}(t_j)$ and $\vec{V}_{\Earth_j}(t_j)$, can be reasonably assumed to be known in the time $t_j$. Now, it is possible to apply a Lambert solver~\cite{izzo2015revisiting} to connect the current gravity-assist in $\vec{R}_{\Earth_j}(t_j)$ with the previous one in $\vec{R}_{\Earth_{j-1}}(t_{j-1})$, by finding the spacecraft velocity $\vec{V}^+_{j-1}$ which the spacecraft leaves the $j-1$ gravity-assist and the incoming velocity $\vec{V}^-_{j}$ for the $j$-th encounter.

With all $\vec{V}^+_{j}$ and $\vec{V}^-_{j}$ found, the incoming and outgoing desired velocities at the infinity, with respect to the gravity-assist body, $\vec{v}_{d\infty}^-$ and $\vec{v}_{d\infty}^+$ respectively, can be calculated in order that the transfers between the gravity-assist bodies are feasible.  

Generally, for the 0SOI-PC, only the $t_j$s are used as decision variables. In this case, and at this point, the procedure would be to find $r_{pj}$ from $\vec{v}_{d\infty}^-$ and to connect the incoming and outgoing leg by an impulse at the periapsis~\cite{izzo2010global}. A collision would be avoided with a non-linear constraint. However, we found trouble in avoiding a collision using this option for the MATLAB built-in optimization routines. As our aim here is not in obtaining an optimal tour design, but only finding a reasonable tour trajectory to analyze the control, we find no problem in simply introducing the $r_{pj}$s as an additional decision variable to avoid the collision.

In our procedure, we assume that the spacecraft will arrive with the desired incoming velocity, $\vec{v}_{\infty}^-=\vec{v}_{d\infty}^-$, and we obtain $v_\infty = ||\vec{v}_\infty^- ||$. Now, we can calculate the half turning angle:

\begin{equation}
\sin \delta = \frac{\mu}{\mu + r_{pj} v_\infty^2},
\end{equation}
and the angular momentum unit vector of the hyperbole:
\begin{equation}
\hat{h} = \frac{\vec{v}_{\infty}^- \times \vec{v}_{d\infty}^+}{||\vec{v}_{\infty}^- \times \vec{v}_{d\infty}^+||},
\end{equation}
to find the actual outgoing velocity at the infinity:
\begin{equation}
\vec{v}_\infty^+ = v_\infty \left[ \cos(2 \delta) \frac{\vec{v}_\infty^-}{v_\infty} + \sin(2 \delta) \frac{\hat{h} \times \vec{v}_\infty^-}{||\hat{h} \times \vec{v}_\infty^-||} \right].
\end{equation}

Finally, an impulse can be imparted to the spacecraft just after leaving the gravity-assist body in order to connect the whole multiple bodies trajectory:
\begin{equation}
\Delta \vec{v} = \vec{v}_{d\infty}^+ - \vec{v}_\infty^+.
\end{equation}

Therefore, in summary, an optimization routine can	 minimize the sum of all the impulses, $|| \sum_{j} \Delta \vec{v} ||$, for the decision variables $t_j$ and $r_{pj}$ in a given sequence $k=1,2,...,j,...,I$ of gravity-assists.

\subsubsection{Patched Conics Tour Design}
\label{sec:patched}

The Zero-SOI assumption works well for interplanetary trajectories, but for planetary systems it shows some limitations~\cite{russell2012survey}. In this case, not only a large mass parameter, as in the case of Earth-Moon~\cite{negri2017studying, negri2019lunar}, affects the accuracy of the 0SOI-PC, but also the assumption of an instantaneous maneuver is not reasonable~\cite{negri2017galilean}. For instance, consider a circular orbit for Io, the small volcanic moon moves on its orbit at an angular rate of ~8.48$^\circ$/h. Therefore, in the tens of minutes inside Io's SOI, the moon moves a considerable amount, directly affecting the calculated swing-by through 0SOI-PC~\cite{negri2017galilean}. For this reason, in this section we not only take into account the magnitude of the SOI in the transfer between the bodies, but also the predicted time inside the SOI, so that the movement of the gravity-assist body in its orbit is taken into account.

In order to accomplish the proposed goal, we have to invert the problem if compared to the 0SOI-PC. Here, it is first completely defined the gravity-assist in each body, the transfer legs between the bodies are calculated and connected after that. Thus, given the encounter time $t_j$, here assumed as the time where the spacecraft is at its periapsis of the hyperbolic trajectory of the $j$-th gravity-assist, and $r_{pj}$, we also add as decision variables: the hyperbolic eccentricity $e$ of the j-th gravity-assist, the argument of periapsis $\omega$, the inclination $i$, and $\Omega$, the longitude of the ascending node (LOAN).

This way, one can find the eccentricity unit vector:
\begin{equation}
\label{eq:evec}
\hat{e} = \begin{bmatrix}
\cos \Omega\cos \omega - \sin \Omega \sin\omega\cos i \\
\sin \Omega \cos \omega + \cos \Omega \sin\omega\cos i \\
\sin\omega \sin i
\end{bmatrix},
\end{equation}
and the angular momentum unit vector:
\begin{equation}
\label{eqn:ang_mom_ver}
\hat{h} = \begin{bmatrix}
\sin i \sin \Omega \\
- \sin i \cos \Omega \\
\cos i
\end{bmatrix}.
\end{equation}
As well as a third unit vector defining the orthogonal system: $\hat{e}_\perp = \hat{h} \times \hat{e}$.

The true anomaly at entering and leaving the SOI is found from the conic equation:
\begin{equation}
\cos \nu = \frac{a(1-e^2)-r_{SOI}}{e r_{SOI}},
\end{equation}
where $a$ is the semi-major axis and found from $r_{pj}$ and $e$, and $r_{SOI}$ is defined as the Laplace sphere of influence:
\begin{equation}
r_{SOI} = R_\Earth \left( \frac{\mu}{\mu_\Sun} \right)^{2/5}.
\end{equation}

Now, it is possible to obtain the position that spacecraft enters and leaves the SOI, respectively as:
\begin{subequations}
\begin{align}
\vec{r}_{SOI}^-&=r_{SOI} [ \cos (-\nu) \hat{e} + \sin (-\nu) \hat{e}_\perp ], \\
\vec{r}_{SOI}^+&=r_{SOI} [ \cos \nu \hat{e} + \sin \nu \hat{e}_\perp ].
\end{align}
\end{subequations}

The respective velocities at the SOI can be found, after calculating the flight-path angle $\cos \gamma = (r_p v_p)/(r_{SOI} v_{SOI})$, as:
\begin{subequations}
\begin{align}
\vec{v}_{SOI}^-=v_{SOI} \left[ \cos \gamma \left( \hat{h} \times \frac{\vec{r}_{SOI}^-}{r_{SOI}} \right)  + \sin \gamma \frac{\vec{r}_{SOI}^-}{r_{SOI}} \right], \\
\vec{v}_{SOI}^+=v_{SOI} \left[ \cos \gamma \left( \hat{h} \times \frac{\vec{r}_{SOI}^+}{r_{SOI}} \right)  + \sin \gamma \frac{\vec{r}_{SOI}^+}{r_{SOI}} \right].
\end{align}
\end{subequations}

Half of the time spent inside the SOI can be easily found after solving the Kepler equation: 
\begin{equation}
\mathcal{ M } = e \sinh E  - E,
\end{equation}
for $\tanh \left( \frac{E}{2} \right) = \sqrt{ \frac{e-1}{e+1}} \tan \left( \frac{\nu}{2} \right) $, as:
\begin{equation}
t_{GA} = \mathcal{ M } \sqrt{ - \frac{a^3}{\mu} }.
\end{equation}

With the time $t_{GA}$ for each hyperbole's leg, one can find the state of the spacecraft relative to the main body before and after the gravity-assist:
\begin{subequations}
\begin{align}
\vec{R}_j^-&= \vec{R}^-_{\Earth_j}(t_j-t_{GA}) + \vec{r}_{SOI}^-, \\
\vec{V}_j^-&= \vec{V}^-_{\Earth_j}(t_j-t_{GA}) + \vec{v}_{SOI}^-, \\
\vec{R}_j^+&= \vec{R}^+_{\Earth_j}(t_j+t_{GA}) + \vec{r}_{SOI}^+, \\
\vec{V}_j^+&= \vec{V}^+_{\Earth_j}(t_j+t_{GA}) + \vec{v}_{SOI}^+.
\end{align}
\end{subequations}

Finally, a Lambert problem is solved to obtain the transfer between all the bodies, by finding the desired incoming velocity $\vec{V}_{dj}^-$ to the j-th gravity-assist and the outgoing desired velocity from the last flyby, $\vec{V}_{dj-1}^+$. This way, an impulse is imparted to the spacecraft just after and before each swing-by in order to make the transfer and to guarantee the chosen conditions for the gravity-assist:

\begin{subequations}
\begin{align}
\Delta \vec{V}_j^- = \vec{V}_{dj}^- - \vec{V}_{j}^-, \\
\Delta \vec{V}_j^+ = \vec{V}_{dj}^+ - \vec{V}_{j}^+.
\end{align}
\end{subequations}

Therefore, given $t_j$, $r_{pj}$, $e_{j}$, $i_{j}$, $\Omega_{j}$, and $\omega_{j}$, as decision variables, the summation of all impulses, $\sum_j \left( ||\Delta \vec{V}_j^-||+||\Delta \vec{V}_j^+|| \right)$, can be minimized to find a tour.

\subsubsection{N-2 Circular Restricted N-Body Problem} 
\label{sec:n-2}

\begin{figure}
	\centering\includegraphics[width=.5\textwidth]{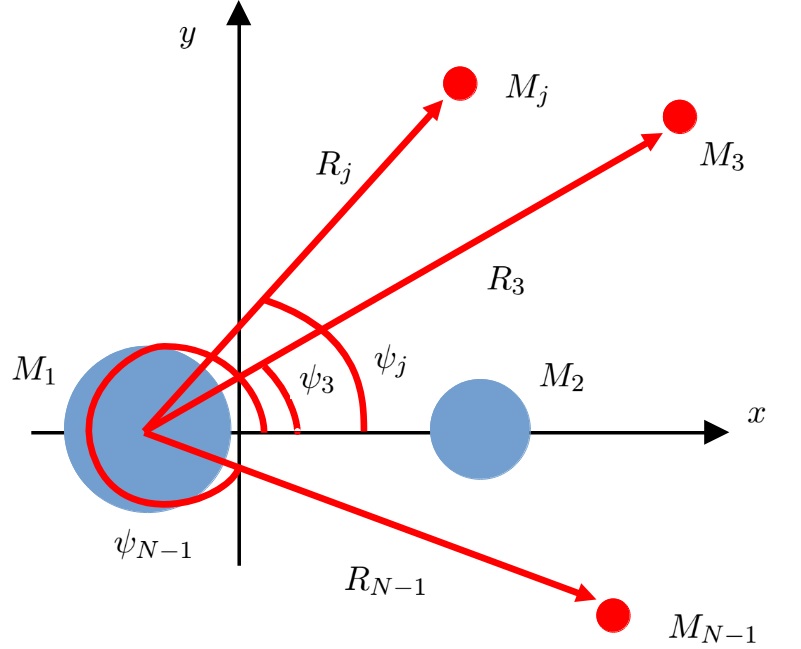}
	\caption{Representation of the N-2 Circular Restricted N-Body Problem, in the synodic frame.}
	\label{fig:N2CRNBP}
\end{figure}	

Even the most precise patched-conics just presented still an approximation of a highly perturbed N-body environment. In an outer planetary system, it is likely the spacecraft will approach many of the massive moons with at least a few SOIs of distance, this renders the trajectory quite chaotic in some cases, driving to very different results than the ones that can be expected with a patched conics approximation.

In order to simulate this N-body problem, we will employ what we are calling an N-2 circular restricted N-body problem (CRNBP). This is an approximation the authors made to simulate a trajectory in an N-body environment, greatly reducing the equations to be integrated. We hope to present a full derivation of these equations soon, in a dedicated paper. However, for now, we just present the equations of motion and refer the reader to Reference~\cite{negri2020generalizing}, where we present the generalization we made for the bicircular restricted four-body problem (BCR4BP) that enabled to obtain the CRNBP. 

Similarly to the BCR4BP, the CRNBP allows to greatly reduce the equations to be integrated. This comes at the expanse of the little physical incoherence of imposing a circular orbit to the bodies (the same is already true for the BCR4BP). Nevertheless, as in the BCR4BP, this is still useful for simpler and general analyses, while retaining a great part of the dynamical complexity.

Figure \ref{fig:N2CRNBP} represents the CRNBP in a synodic frame that is rotating with the primaries of mass $M_1$ and $M_2$, respectively, which are assumed to describe a circular orbit. The other bodies, $M_j$, $j=3,...,N-1$, are also assumed to have a circular orbit about $M_1$, coplanar to the one of $M_1$ and $M_2$, and totally defined by the angle $\psi_j$. Therefore, the equations of motion of a small body moving in this frame, in canonical units, are:

\begin{subequations}
\label{eq:N2CRNBP}
\begin{align}
\begin{split}
\ddot{x} &= 2 \dot{y} + x - \frac{\mu_1}{r_1^3}(x+\mu_2) - \frac{\mu_2}{r_2^3} (x-\mu_1) - \sum_{j=3}^{N-1} \mu_j \left[ \frac{1}{r_j^3} (x+\mu_2-R_j \cos \psi_j ) + \right. \\ & \left.  \sum_{k=1,k\neq j}^{N-1} \frac{\mu_k}{(R_k^2+R_j^2-2 R_k R_j \cos(\psi_k-\psi_j) )^{3/2}} (R_j \cos \psi_j - R_k \cos \psi_k ) \right], 
\end{split} \\
\begin{split}
\ddot{y} &= - 2 \dot{x} + y - \frac{\mu_1}{r_1^3} y - \frac{\mu_2}{r_2^3} y - \sum_{j=3}^{N-1} \mu_j \left[ \frac{1}{r_j^3} (y-R_j \sin \psi_j ) + \right. \\ & \left.  \sum_{k=1,k\neq j}^{N-1} \frac{\mu_k}{(R_k^2+R_j^2-2 R_k R_j \cos(\psi_k-\psi_j) )^{3/2}} (R_j \sin \psi_j - R_k \sin \psi_k ) \right], 
\end{split} \\
\ddot{z} &= - \frac{\mu_1}{r_1^3} z - \frac{\mu_2}{r_2^3} z - \sum_{j=3}^{N-1}  \frac{\mu_j}{r_j^3} z. 
\end{align}
\end{subequations}
The mass parameter of each body is defined as $\mu_j=M_j/(M_1+M_2)$, and $R_j$ are the distance of each body to $M_1$. The $\psi_j$ can be solved in time by the analytical expression:
\begin{equation}
\psi_j = \psi_{0j}+ (n_j-n_{12}) t,
\end{equation}
where $\psi_{0j}$ represents the initial angle, $n_j$ is the mean motion of the $j$-th in canonical units and $n_{12}$ is the mean motion of $M_1$ and $M_2$ about their center of mass, which is $n_{12}=1$ in canonical units.

Note that this description of the problem reduces the 6N first-order differential equations of a N-body problem to simply six differential equations and N-2 analytical expressions.

\section{Robust Keplerian Path-Following Control}

In the hyperbolic approach to the gravity-assist body, the most important aspect to be guaranteed is the geometry of the trajectory. The small third-body perturbation, drag, and other disturbances effects that could be present in such a scenario, are very unlikely to, together with a control to cancel they out, affect the trajectory in a way to delay or advance the spacecraft in tens of minutes or hours. Moreover, a possible accumulated delay or advance caused by consecutive gravity-assists are most likely to be dealt with by the guidance algorithm (i.e., recalculating the tour trajectory) rather than by control enforcement. Therefore, for the hyperbolic trajectory, a path following control is much more suitable than a reference tracking. In this way, we apply the robust Keplerian path-following control (RKPFC) derived by Reference~\cite{negri2020novel}, which showed promising results for small body missions in terms of fuel savings and operational requisites~\cite{negri2021autonomous}.

The RKPFC is a sliding-mode control, but it uses a different set of sliding surfaces than the ones usually applied. The sliding surface $\vec{s}$ is defined in terms of the integrals of motion of the two-body problem. Once the equilibrium condition for the sliding surface is reached, $\vec{s} = 0$, the controlled spacecraft asymptotically converges to the desired Keplerian geometry. The sliding surface is~\cite{negri2020novel}:

\begin{equation}
\label{eqn:sliding_surface}
\vec{s} = \begin{bmatrix}
\tilde{\vec{e}} \cdot (\lambda_R \hat{r} + \hat{\theta}) \\
\tilde{h} \\
\hat{h}_d \cdot (\lambda_N \hat{r} + \hat{\theta})
\end{bmatrix}=0,
\end{equation}
where $\lambda_R>0$ and $\lambda_N>0$ determine the rate of convergence to the desired Keplerian geometry, as shown in Proposition 1 in Reference~\cite{negri2020novel}, $\tilde{\vec{e}} = \vec{e}-\vec{e}_d$ represents the error between, respectively, the current and desired eccentricity vectors, $\hat{h}_d$ is the desired specific angular momentum unit vector and $\tilde{h}=h-h_d$ is the error in the magnitude of the specific angular momentum. The unit vectors $\hat{r}$ and $\hat{\theta}$ are the unit vectors of the radial-transverse-normal (RTN) coordinates, $\hat{r}=\vec{r}/||\vec{r}||$ and $\hat{\theta}=\hat{h}\times \hat{r}$.

The desired unit vector $\hat{h}_d$, defining the orbital plane, can be obtained from Eq. (\ref{eqn:ang_mom_ver}) by choosing a desired inclination $i_d$ and LOAN $\Omega_d$. The current and desired angular momentum are obtained as $h=|| \vec{r} \times \vec{v} ||$ and $h_d=\sqrt{\mu a_d (1-e_d^2)}$, respectively. Finally, the current and desired eccentricity vectors are respectively: $\vec{e}=\frac{1}{\mu} (\vec{v} \times \vec{h} - \mu \hat{r})$ and $\vec{e}_d=e_d \hat{e}_d$, with $\hat{e}_d$ found from Eq. (\ref{eq:evec}) for $\Omega_d$, $\omega_d$ and $i_d$.

Using the sliding surface in Eq. (\ref{eqn:sliding_surface}), robustness to bounded disturbances and asymptotic convergence to the geometry of a Keplerian orbit can be obtained using the control:
\begin{equation}
\label{eq:control_theo2}
\vec{u}  = - [RTN]^{-1} F^{-1} ( G + K \text{sgn}(\vec{s})) - \vec{f},
\end{equation}
$K \in \mathbb{R}^{3\times 3}$ is a diagonal positive definite matrix, the function $\text{sgn}(\vec{s}) \in \mathbb{R}^{3 \times 1}$ represents the sign function taken in each component of $\vec{s}$, the matrices $F$ and $G$ are defined by:

\begin{subequations}
\begin{align}
F &= \frac{1}{h\mu} \begin{bmatrix}
-h^2 & \left[2\lambda_R h-(\vec{v}\cdot\hat{r})r\right]h & -\mu r (\vec{e}_d\cdot\hat{h}) \\
0 & \mu rh & 0 \\
0 & 0& \mu r (\hat{h}_d\cdot\hat{h}) 
\end{bmatrix}, \\
G &=\frac{h}{r^2} \begin{bmatrix}
\tilde{\vec{e}}\cdot(\lambda_R\hat{\theta}-\hat{r}) -1 \\
0 \\
\hat{h}_d\cdot(\lambda_N\hat{\theta}-\hat{r})
\end{bmatrix},
\end{align}
\end{subequations}
and the matrix $[RTN]$ is a matrix that transforms from the Cartesian coordinates to RTN, defined as:

\begin{equation}
[RTN] = \begin{bmatrix}
\hat{r}^\mathbb{T} \\ \hat{\theta}^\mathbb{T} \\ \hat{h}^\mathbb{T}
\end{bmatrix}
\end{equation}
with the superscript $\mathbb{T}$ representing the transpose.

The control in Eq. (\ref{eq:control_theo2}) lies on the assumption that the magnitude of $\vec{h}$ and $\vec{r}$ are not zero. And, if defined an angle $\beta$ such that $\cos \beta = \hat{h} \cdot \hat{h}_d$, this angle is bounded by $\beta<90^\circ$. Although this assumption is generally of no to little harm for the orbit keeping problem~\cite{negri2021autonomous}, it can cause issues for the gravity-assist control. We will deal with this point later.

The diagonal gain matrix $K$ can be chosen to guarantee convergence for unknown bounded disturbances $\lvert d_R \rvert < D_R$, $\lvert d_T \rvert < D_T$, and $\lvert d_N \rvert < D_N$, for $\vec{d}_{RTN}= \begin{bmatrix}
d_R & d_T & d_N
\end{bmatrix}^\mathbb{T}=[RTN]\vec{d}$, as \cite{negri2020novel}:

\begin{subequations}
\label{eqn:K_matrix}
\begin{align}
K_{1,1} &\geq  \frac{h}{\mu} D_R + \left\lvert \frac{2\lambda_R h-(\vec{v}\cdot\hat{r})r}{\mu} \right\rvert D_T + \frac{ r \left\lvert \vec{e}_d\cdot\hat{h}\right\rvert}{h}  D_N, \\
K_{2,2} &\geq r D_T, \\
K_{3,3} &\geq r \frac{\hat{h}_d\cdot\hat{h}}{h} D_N,
\end{align}
\end{subequations}
in which $K_{j,j}$, $j=1,2,3$, are the diagonal elements of the matrix $K$. 

In order to make easier the analysis and simulations, we can approximate the control in Eq. (\ref{eq:control_theo2}) by an impulse:
\begin{equation}
\Delta \vec{V} = \Delta t \vec{u} ,
\end{equation}
where $\Delta t$ is the control update time step.

\subsection{Practical considerations}

The discontinuity in the function $\text{sgn}(\vec{s})$ in the sliding-mode control is known to cause chattering in many applications~\cite{slotine1991applied}. However, this can be easily circumvent by substituting it by a continuous function, with a little trade in performance. Here, we substitute the sign function by a saturation function $\text{sat}(\vec{s},\vec{\Phi})$, defined as:
\begin{equation}
\text{sat}(s_j,\Phi_j) = \begin{cases} 1, &\text{$s_j>\Phi_j$} \\\frac{s_j}{\Phi_j}, &\text{$-\Phi_j\leq s_j\leq \Phi_j$} \\ -1, &\text{$s_j<-\Phi_j$} \end{cases},
\end{equation}
for $j=1,2,3$ representing each component of $\vec{s}$ and $\vec{\Phi}$.

As discussed in Reference~\cite{negri2021autonomous}, a great advantage of the RKPFC is its ability to easily accommodate periods with the thrusters turned off. If they are turned back on after a long period of idling, there is no concern for which reference to choose and if it can cause an adverse behavior or consequence (e.g., a reference too far posing the spacecraft in risk of collision or in unnecessary waste of fuel), as the RKPFC will only recover the Keplerian geometry. 

In this work we will propose an alternative of control switch inspired in the Schmitt trigger as follows:
\begin{equation}
\text{hys}(\vec{\chi}) = \begin{cases} 1, &\text{if any } \text{$\chi_j>\chi_j^+$} \\ 0, &\text{if all } \text{$\chi_j<\chi_j^-$} \\ 1, &\text{if $\chi_j^- \leq \chi_j \leq \chi_j^+$ and hys$_p(\vec{\chi})=1$} \\ 0, &\text{if $\chi_j^- \leq \chi_j \leq \chi_j^+$ and hys$_p(\vec{\chi})=0$} \end{cases},
\end{equation}
j=1,...,5, in which $\text{hys$_p$}$ is the value of $\text{hys}$ in the previous iteration and $\vec{\chi}_j$ is:
\begin{equation}
\vec{\chi} = \begin{bmatrix}
|a - a_d| \\ |e - e_d| \\ |i - i_d| \\ |\omega - \omega_d| \\ |\Omega - \Omega_d| 
\end{bmatrix}.
\end{equation}
The vector $\vec{\chi}^-$ and $\vec{\chi}^+$ represent the lower and upper bounds for the hysteris in each component of $\vec{\chi}$.

\subsection{B-Plane Linear Quadratic Regulator}

As mentioned earlier, the RKPFC lies on the assumption that for $\cos \beta = \hat{h} \cdot \hat{h}_d$, this angle is bounded as $\beta<90^\circ$. However, for a gravity-assist, this condition is not guaranteed to hold. In fact, it might occur with some frequency, depending on the accuracy of the guidance algorithm to predict the spacecraft state upon arriving at the gravity-assist body. To address this issue, we propose a solution. Instead of relying on a reference tracking strategy, we employ an infinite horizon linear quadratic regulator (LQR) to regulate the impact parameter vector of the approaching spacecraft.

\begin{figure}[htb!]
	\centering\includegraphics[width=.5\textwidth]{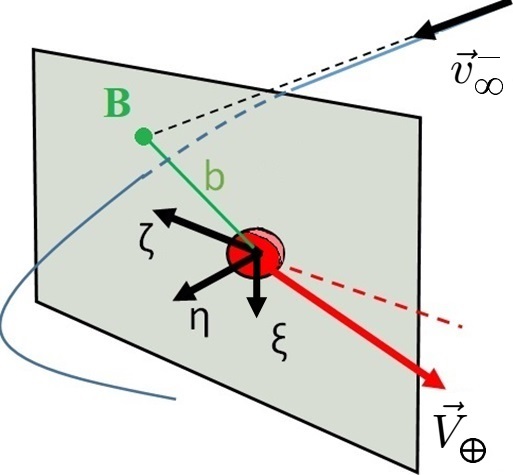}
	\caption{Representation of the b-plane.}
	\label{fig:bplane}
\end{figure}

Here, we use the b-plane, represented in Figure \ref{fig:bplane}, which is a plane extensively applied for gravity-assist design. It is defined as perpendicular to the velocity at the infinity ($\vec{v}_\infty^-$), and containing the gravity-assist body center. With this definition, it is possible to apply an orthogonal frame centered in the gravity-assist body and defined as:
\begin{subequations}
\begin{align}
\hat{\eta} &= \frac{\vec{v}_\infty^-}{v_\infty}, \\
\hat{\xi} &= \frac{\vec{V}_\Earth \times \hat{\eta}}{||\vec{V}_\Earth \times \hat{\eta}||}, \\
\hat{\zeta} &= \hat{\xi} \times \hat{\eta}.
\end{align}
\end{subequations}

The impact parameter is a vector represented in the b-plane that indicates the point where the velocity at the infinity pierces the plane. If the spacecraft is far from the gravity-assist body, i.e. roughly outside its SOI, the impact parameter can be easily described on the b-plane as:
\begin{equation}
\label{eq:imp_par}
\vec{b} = J \vec{r},
\end{equation}
in which the matrix $J$ is:
\begin{equation}
J = \begin{bmatrix}
\hat{\zeta}^\mathbb{T} \\ \hat{\xi}^\mathbb{T}
\end{bmatrix}.
\end{equation}

The desired impact parameter can be found using two-body problem relations as:
\begin{equation}
\vec{b}_d = J a_d \left( \frac{\sqrt{e_d^2 - 1}}{e_d} \hat{e}_{d\perp} - \frac{e_d^2-1}{e_d} \hat{e}_d \right)
\end{equation}

Defining $\vec{X} = \begin{bmatrix} \vec{b}-\vec{b}_d &\ \dot{\vec{b}} \end{bmatrix}^\mathbb{T}$, considering that $\frac{d}{dt} ( J ) \approx 0$ and $\vec{w}=\vec{u}+\vec{f}$, and assuming that the function $\vec{f}$ is nearly constant, the system can be described as a linear time invariant system:
\begin{equation}
\dot{\vec{X}} = A \vec{X} + B \vec{w},
\end{equation}
for the matrices $A$ and $B$ respectively defined as:
\begin{subequations}
\begin{align}
A &= \begin{bmatrix} 0_{2\times 2} & I_{2 \times 2} \\ 0_{2\times 2} & 0_{2\times 2}  \end{bmatrix} ,\\
B &= \begin{bmatrix} 0_{2\times 3} \\ J \end{bmatrix} .
\end{align}
\end{subequations}

Therefore, an infinite horizon LQR controller can be easily obtained as~\cite{bryson1975applied}:
\begin{equation}
\vec{u} = - K \vec{X} - \vec{f},
\end{equation}
where $K$ is given by:
\begin{equation}
K = R^{-1} B^\mathbb{T} P,
\end{equation}
with $P$ being the solution for the Riccati equation:
\begin{equation}
A^\mathbb{T} P + P A - PBR^{-1}B^\mathbb{T} P + Q = 0,
\end{equation}
considering the cost function:
\begin{equation}
\mathcal{J} = \int_0^\infty (\vec{X}^\mathbb{T}Q \vec{X} + \vec{u}^\mathbb{T}R \vec{u}) dt.
\end{equation}

\section{Analysis and Discussion}

In April 2017, Cassini made its 126th and last flyby of Titan, ultimately leading it to the disintegration in Saturn's atmosphere, in September 2017, for satisfying planetary protection requirements~\cite{bellerose2018cassini}. Among the more of a hundred Titan's flybys, the closest approach occurred in June 2010, when the spacecraft had its encounter periapsis at 880 km altitude~\cite{waite2013model}. At this altitude, Titan's atmosphere is of little impact on the spacecraft trajectory, and small corrective burns in the order of cm/s would compensate for its effects~\cite{pelletier2006atmospheric}. As the altitude decreases, its effects grow exponentially, rapidly reaching the same order of magnitude of the Titan's gravity. It is unlikely that a spacecraft is brought under such severe conditions with no atmospheric modelling and preparation to deal with it. However, under the autonomous gravity-assist, a somewhat challenging unexpected environment could not be discarded. In fact, for the Cassini mission itself, it is reported the unexpected variations in density found in the firsts Titan's gravity-assists, concerning the engineers and leading to a reassessment of the minimum safety altitude that was originally set at 950 km altitude at that time~\cite{pelletier2006atmospheric}.

\begin{figure}[!htb]
\centering
\subfloat[Trajectories]{\includegraphics[width=.5\textwidth]{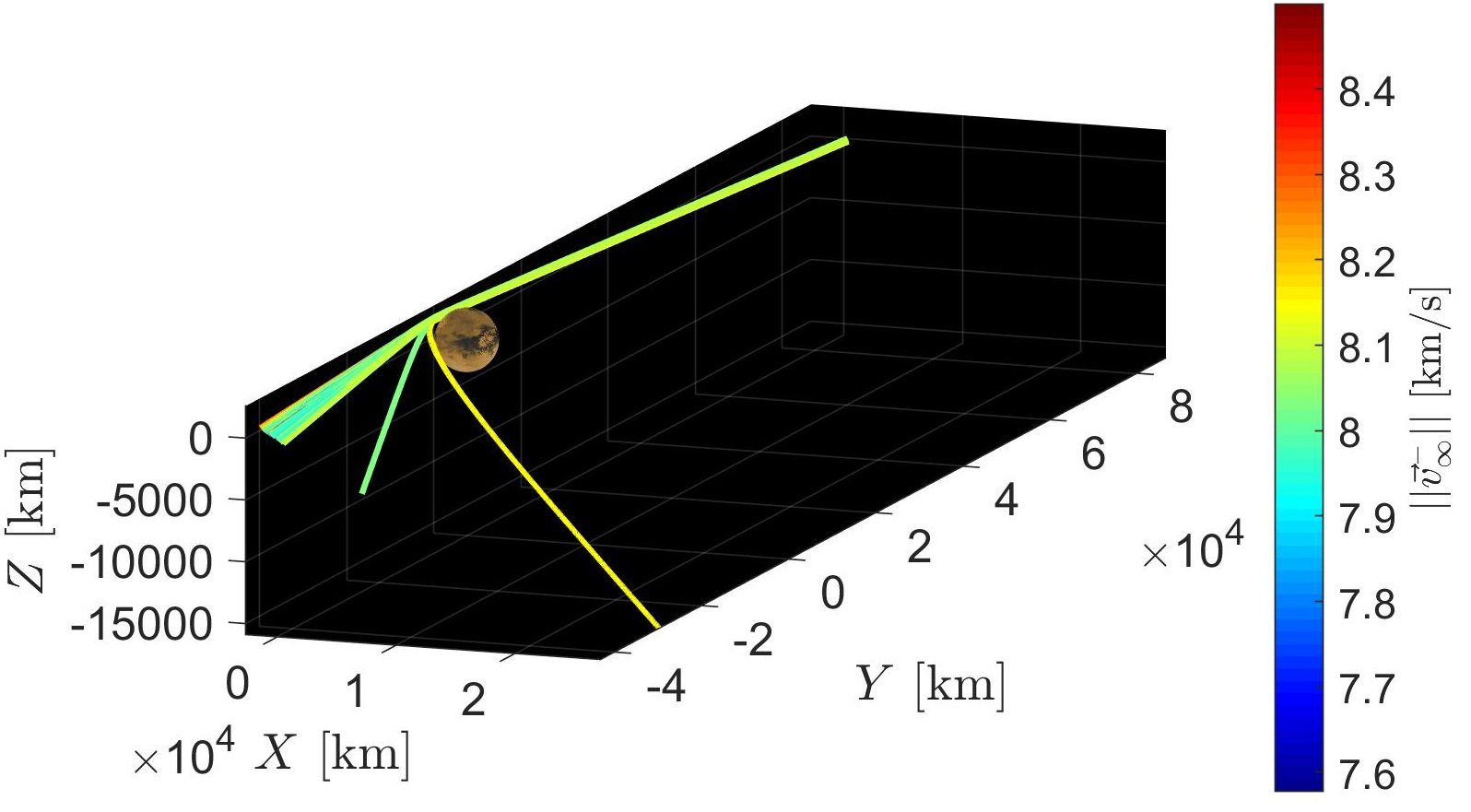}\label{fig:titan_un_3D}}\\
\subfloat[$a$ and $e$]{\includegraphics[width=.5\textwidth]{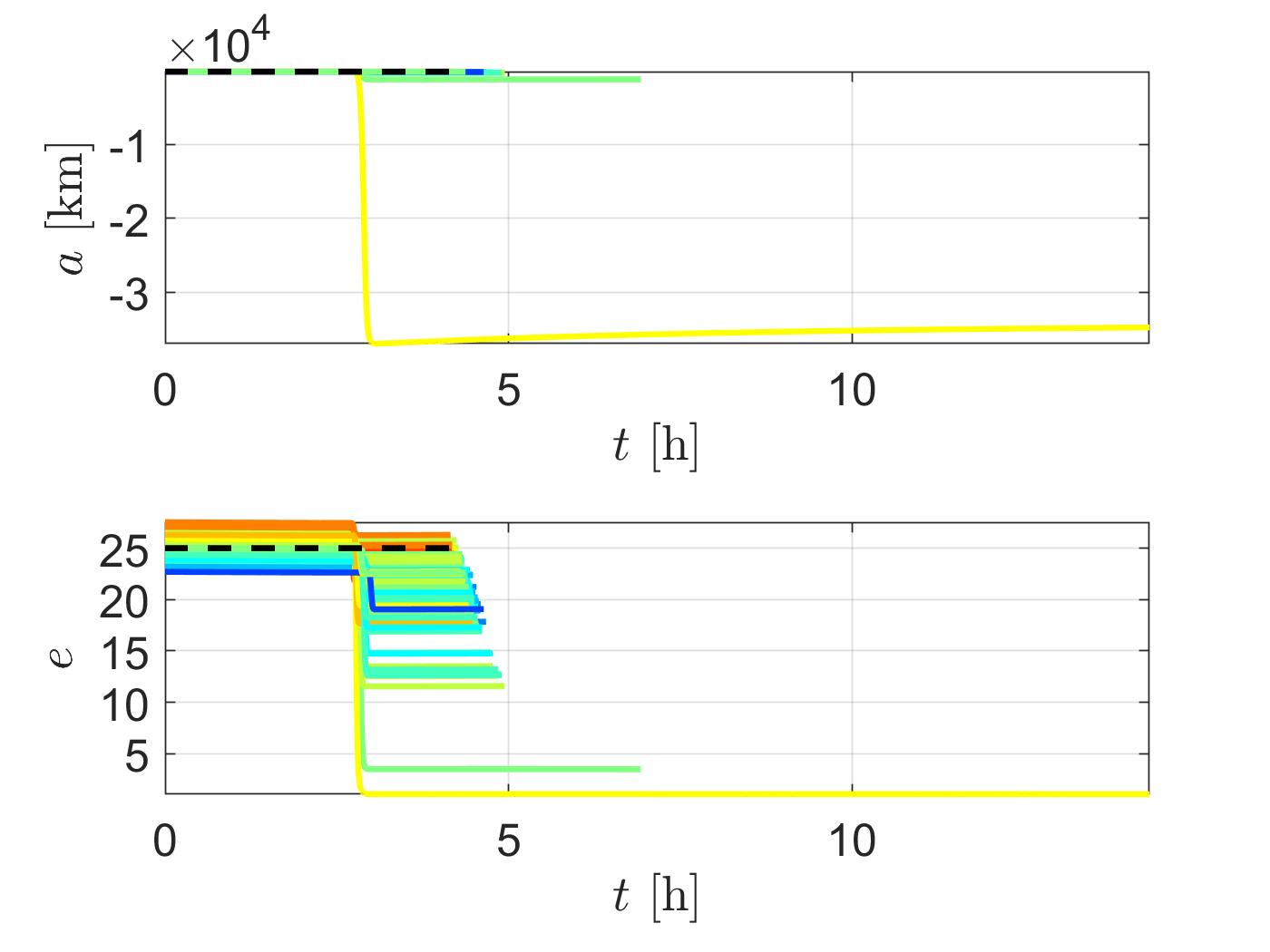}\label{fig:titan_un_ae}}  
\subfloat[$i$, $\omega$, and $\Omega$]{\includegraphics[width=.5\textwidth]{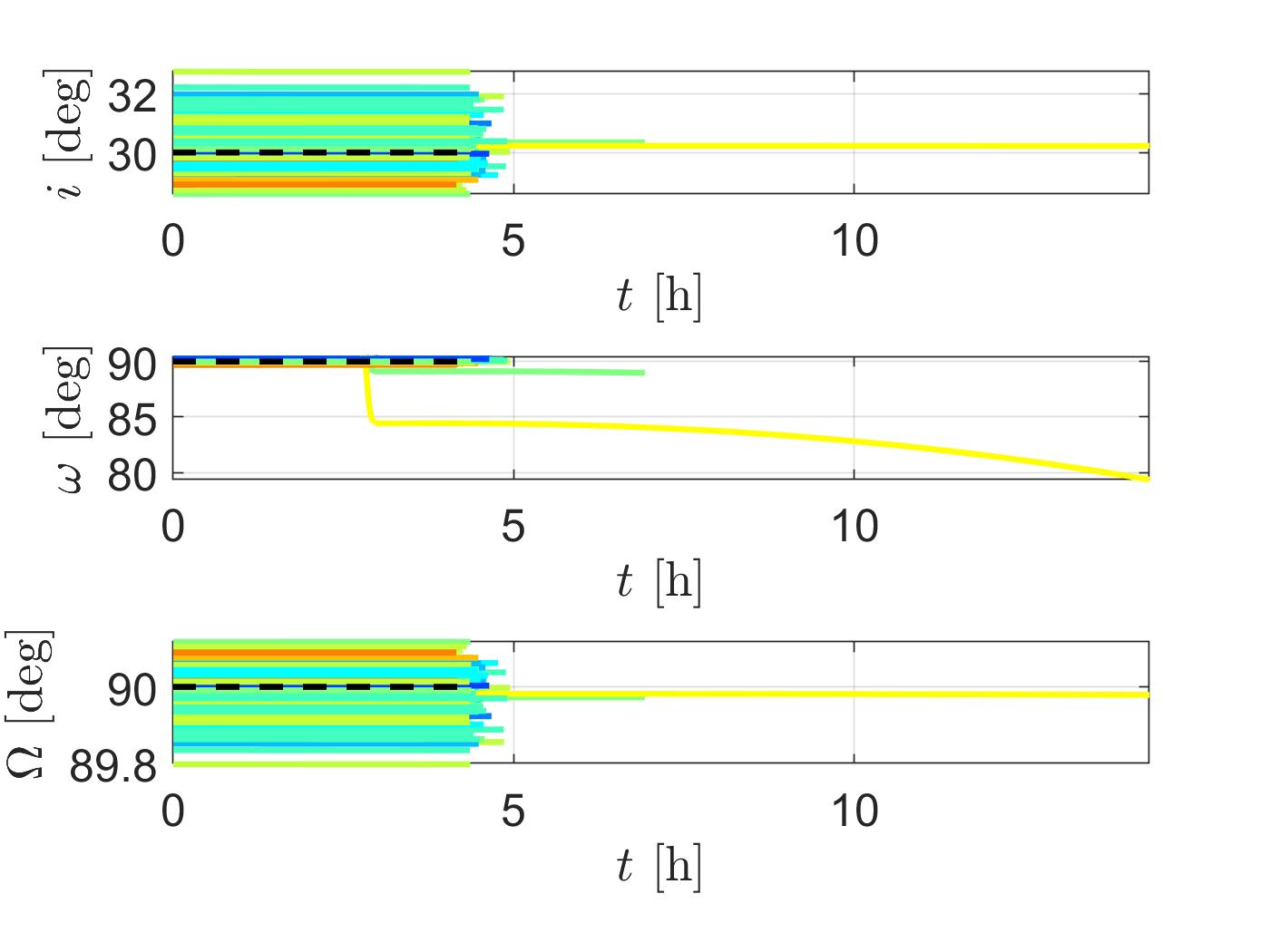}\label{fig:titan_un_iwO}}

\caption{Monte Carlo of the uncontrolled Titan's gravity-assist.}
\label{fig:titan_un}
\end{figure}

\begin{figure}[!htb]
\centering
\subfloat[Trajectories]{\includegraphics[width=.5\textwidth]{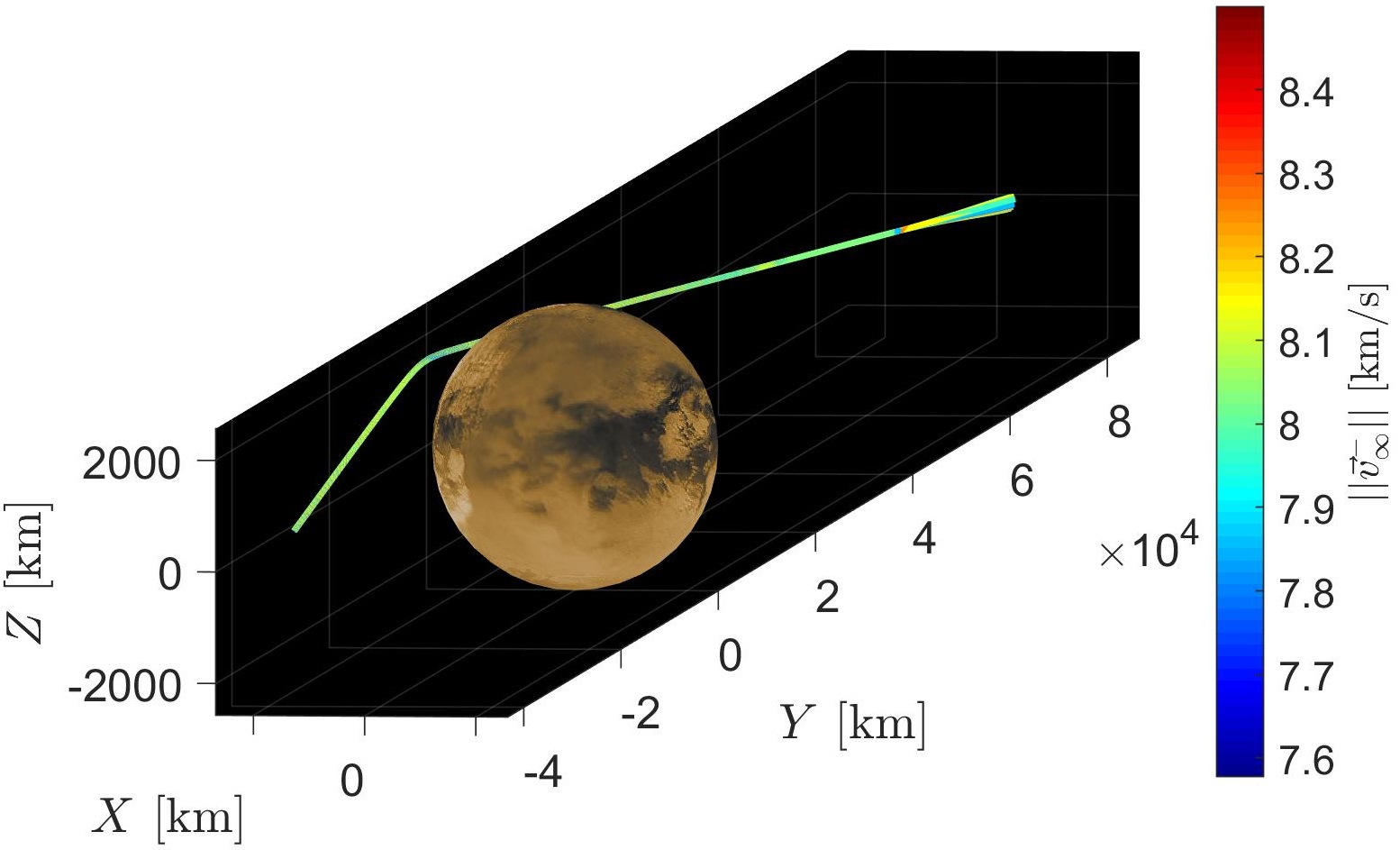}\label{fig:titan_con_3D}}
\subfloat[$\sum \Delta V$]{\includegraphics[width=.5\textwidth]{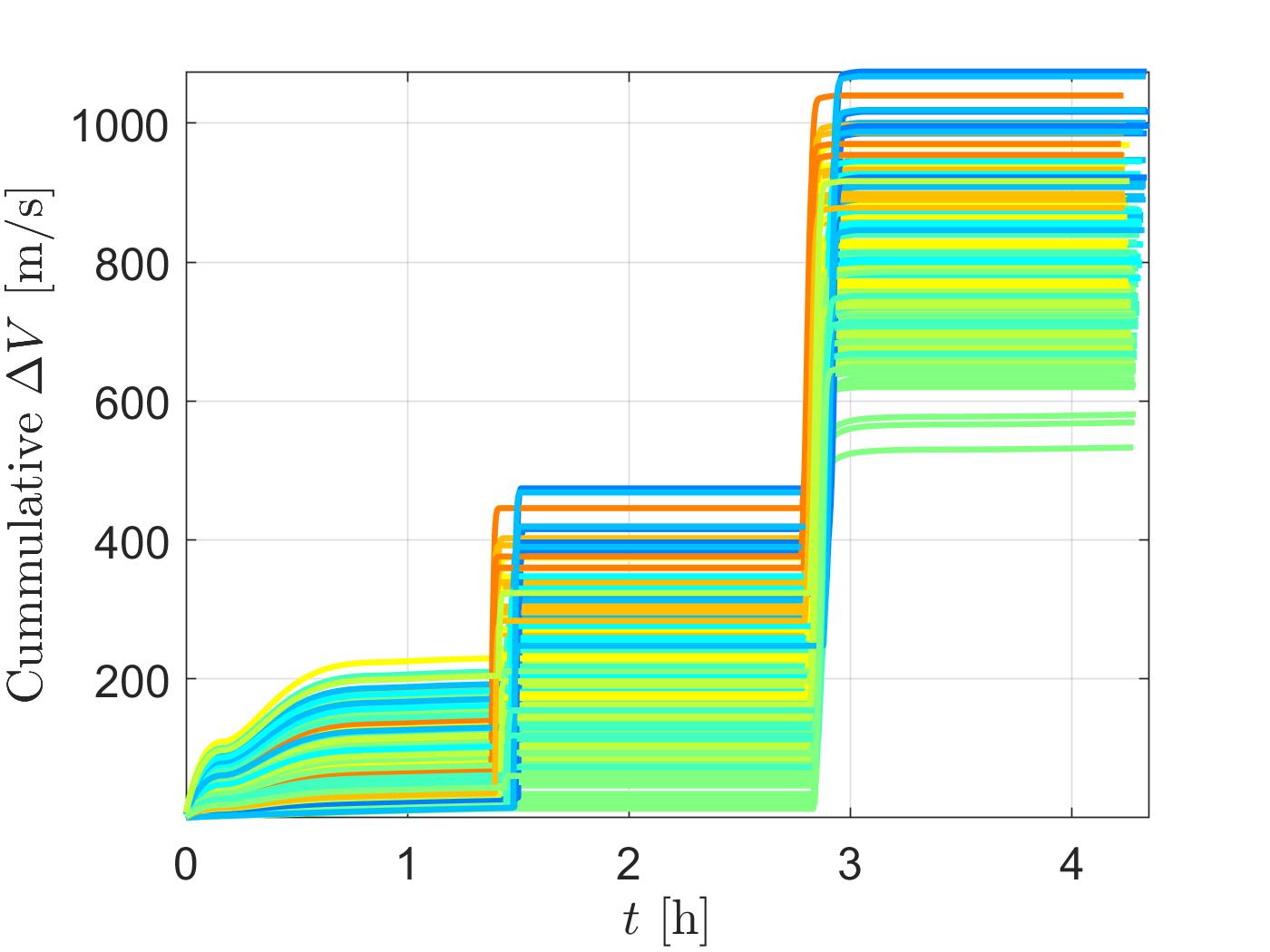}\label{fig:titan_con_dV}}  \\
\subfloat[$a$ and $e$]{\includegraphics[width=.5\textwidth]{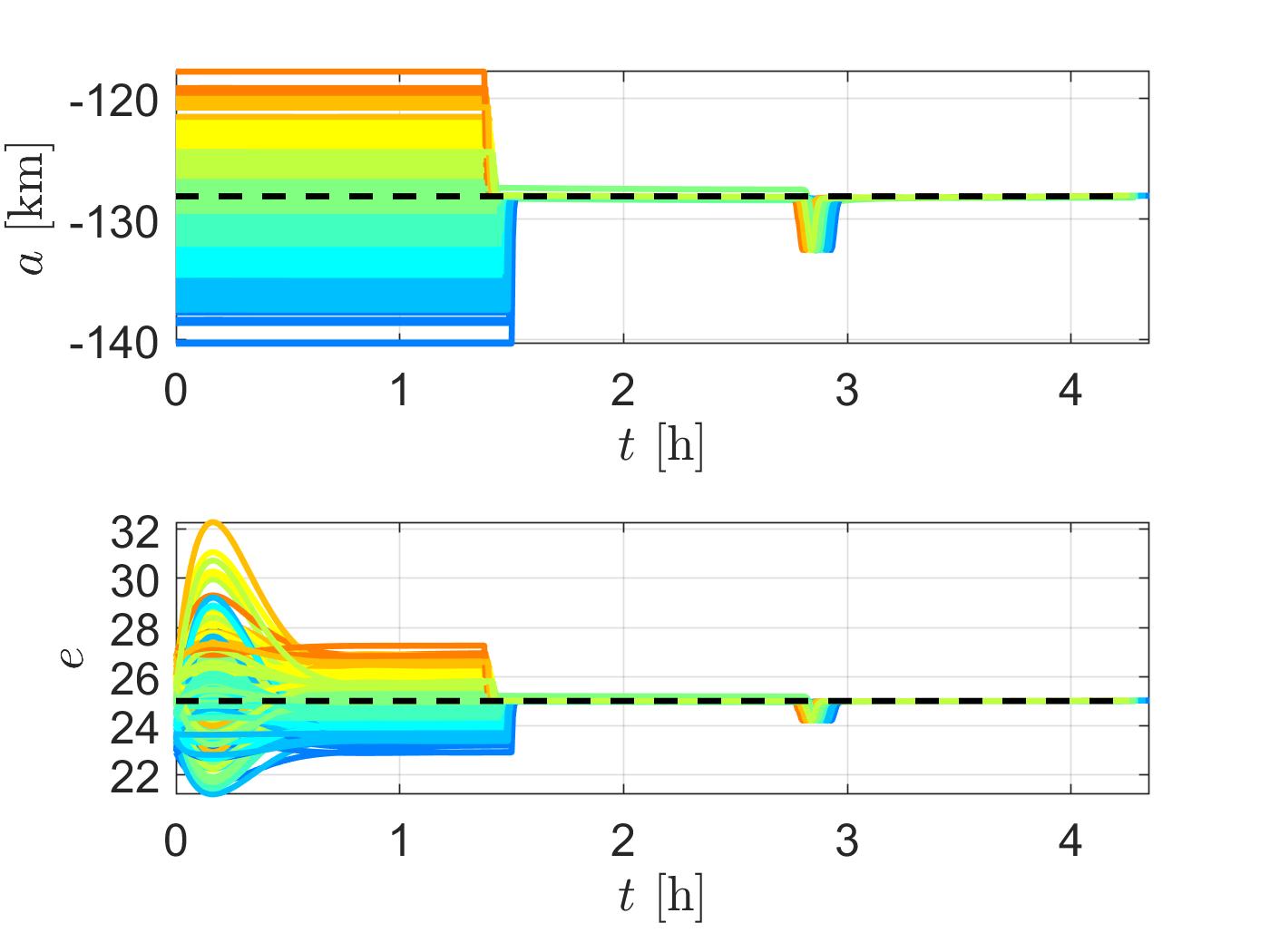}\label{fig:titan_con_ae}}  
\subfloat[$i$, $\omega$, and $\Omega$]{\includegraphics[width=.5\textwidth]{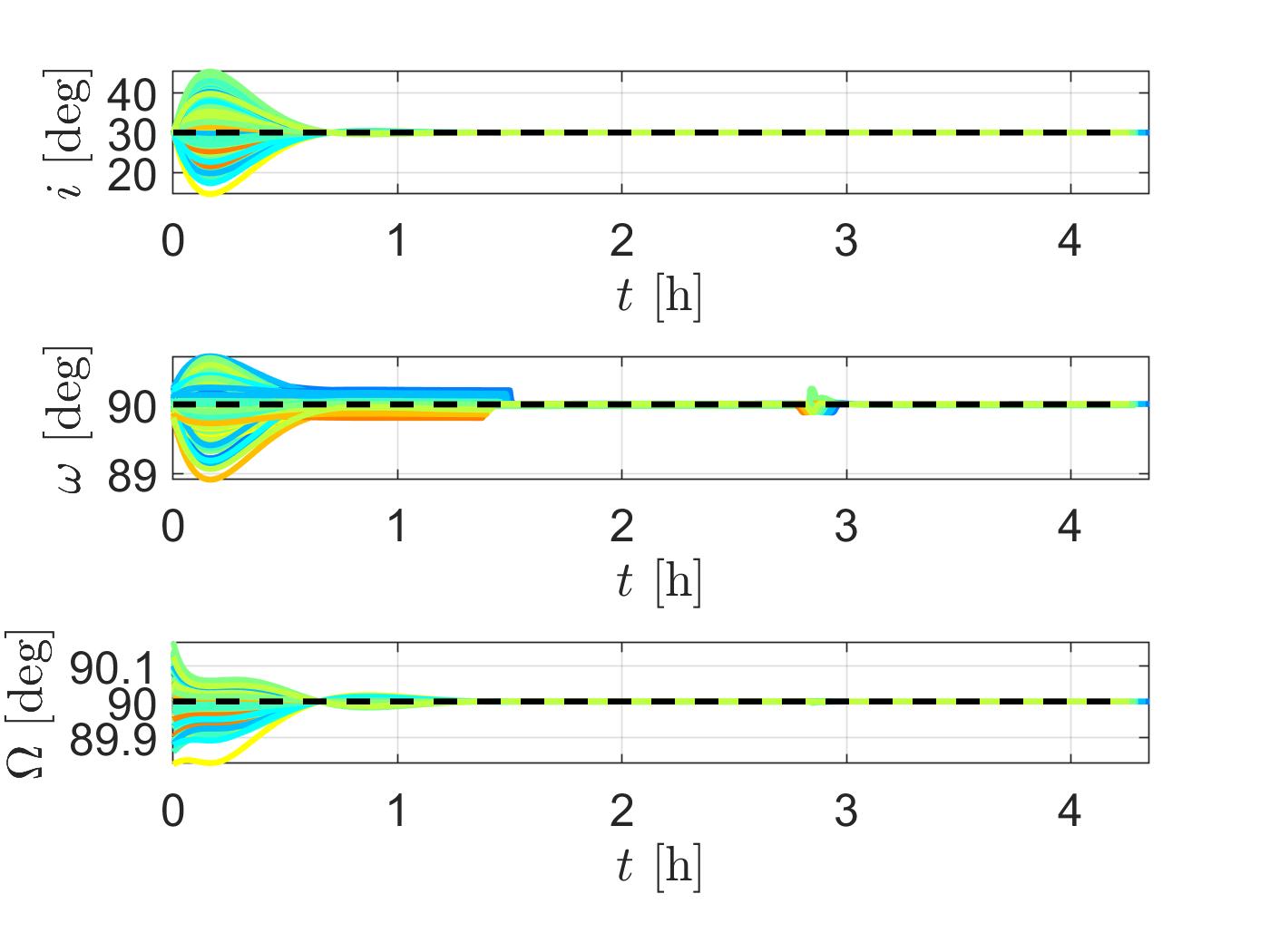}\label{fig:titan_con_iwO}}

\caption{Monte Carlo of the controlled Titan's gravity-assist.}
\label{fig:titan_con}
\end{figure}

Given this context, we choose Titan as the first simulation scenario for our control. We assume the following desired orbital elements for the gravity-assist: $e_d=25$, $i_d=30^\circ$, $\omega_d=90^\circ$, and $\Omega_d=90^\circ$, with a desired semi-major axis $a_d=-128.1$ km for a periapsis altitude of 500 km. We run simulations with the spacecraft approaching Titan at 3 SOIs from its center, with the position and velocity solved for the corresponding desired orbital elements, integrated until the spacecraft leaves Titan's SOI. Monte Carlo simulations are conducted with 200 samples, considering a normal distribution of error for the initial states. The 3D encounter errors in the last 10 Titan flybys by Cassini were within 3 km~\cite{bellerose2018cassini}. However, as an autonomous gravity-assist involves diverse forms of guidance algorithms for transferring the spacecraft between multiple bodies (e.g., an embedded simplified approximation, embedded complex and precise dynamical model, or an uploaded batch of highly precise guidance calculated by ground orbit determination), we consider a higher magnitude of error for the encounter. We assume that the magnitude of the impact parameter, Eq. (\ref{eq:imp_par}), has a 1-$\sigma$ dispersion of 50 km, while the velocity has an error distribution of 2\% in the magnitude of the desired velocity.

 It is considered as disturbances the gravitational acceleration of Saturn, Eq. (\ref{eq:d3B}), and the drag force of the Titan's atmosphere, Eq. (\ref{eq:dD}), when the spacecraft is within 1,500 km altitude. The atmosphere density profile is obtained from an exponential rough fit with the figure 11 in Reference~\cite{waite2013model}, which is good enough for our purposes, as follows:
\begin{equation}
\rho(h) = \exp \left[ \Theta \exp \left( \Xi h \right) + \Lambda \exp \left( \Pi h \right) \right],
\end{equation}
where $h$ is the altitude, and $\Theta=-19.0254$, $\Lambda=17.5748$, $\Xi=3.0747\times10^{-7}$, and $\Pi=-1.2258\times10^{-6}$.

Figure \ref{fig:titan_un} shows the results obtained if no control authority is considered. Note in Figure \ref{fig:titan_un_3D} that the dispersion of the outgoing hyperbole leg is so large that is even visually perceptive. The two out-layer trajectories easily identified in Figures \ref{fig:titan_un_ae} and \ref{fig:titan_un_iwO} are the result of special interaction conditions with Titan's atmosphere, reducing the trajectory eccentricity by more than 10, while the other trajectories are within 10. The mean and standard deviation of the outgoing orbital elements are presented in Table~\ref{tab:label}, as Titan unc. 500.

In Figure \ref{fig:titan_con} are presented the results assuming the control command. We assume that in the incoming leg outside the SOI, the LQRC is working to bring the spacecraft at least close to pierce the b-plane in the right position. After that, the RKPFC assumes as the control input, up to reach the SOI in the outgoing leg. In the LQR, the Q and R are assumed as the diagonal matrices: $\text{diag}\left(\begin{bmatrix} 10^{-6} & 10^{-6} & 10^{-2} & 10^{-2} \end{bmatrix}\right)$ and $\text{diag}\left(\begin{bmatrix} 10^{5} & 10^{5} & 10^{5} \end{bmatrix}\right)$. For the RKPF, the matrix K is continuously calculated using Eqs. \ref{eqn:K_matrix}, for $D_R=D_T=D_N=50$ m/s$^2$, with $\lambda_R=\lambda_N=2$. It is also assumed that each element of $\vec{\Phi}$ is 50 times the corresponding element in K, and boundaries for the control switch are: $\vec{\chi}^-= \begin{bmatrix} 100 & 0.5 & 0.1 & 0.1 & 0.1 \end{bmatrix}$ and $\vec{\chi}^+= \begin{bmatrix} 1000 & 1.0 & 0.5 & 0.5 & 0.5 \end{bmatrix}$ (in meters or degrees). It is also assumed a control update time of 20 seconds.

One can distinguish in Figure \ref{fig:titan_con} three clear distinct phases. In the first one, the LQR controls the position of the spacecraft in the b-plane, up to 1-2 h. After that, a large spike in the budget $\Delta V$, Figure \ref{fig:titan_con_dV}, is the indicative that the spacecraft entered Titan's SOI and had to change its velocity, as there is no control for velocity in the $\hat{\eta}$ direction of the LQR control. A second spike occurs at about 3 h, when the spacecraft is inside Titan's atmosphere trying to compensate the drag force. When within Titan's atmosphere, it is noted little bumps in the semi-major axis and eccentricity, Figure \ref{fig:titan_con_ae}. In order to completely remove these bumps, the gain matrix $K$ should be much larger, or $\vec{\Phi}$ much lower, what would make the control effort much less efficient in other parts of the trajectory. Nevertheless, these bumps present a much lower magnitude than the observed for the uncontrolled case, e.g., the eccentricity is within 1, much lower if compared to the order of 10 of the uncontrolled simulation. It should also be considered that the drag disturbance reach levels of 1 order of magnitude larger than the gravitational term. In a real scenario, if such close proximity to Titan is made, it very likely that a good enough drag model would be at disposal, with the control needing to compensate only small deviations. In this case, a much lower gain matrix and $\vec{\Phi}$ could be chosen, improving the overall performance. The mean and standard deviation of the outgoing parameters are presented in Table~\ref{tab:label} as Titan con. 500. The mean budget $\Delta V$ for this case is of 783.7 m/s, with a standard deviation of 106.7 m/s. Assuming a much more reasonable periapsis altitude of 750 km, and for $D_R=D_T=D_N=0.01$ m/s$^2$, $\lambda_R=\lambda_N=1.5$ and $\Phi$ as 15 times each corresponding element of $K$, one can note in Table~\ref{tab:label} that the control retain the same performance presented before (case Titan con. 750) with a mean budget $\Delta V$ of 212.7 m/s, with a standard deviation of 100.9 m/s.

\begin{table}[htbp]
	\fontsize{9}{9}\selectfont
    \caption{Monte Carlo simulation for each case, magnitudes of the mean (1-$\sigma$) at leaving the body's SOI.}
   \label{tab:label}
        \centering 
   \begin{tabular}{| p{.08\textwidth} | p{.12\textwidth} | p{.12\textwidth} | p{.12\textwidth} | p{.12\textwidth} | p{.12\textwidth} | p{.13\textwidth}  | } % Column formatting, 
      \hline 
      Cases  & $a$ [km] & $e$  &  $i$ [$^\circ$]   & $\omega$ [$^\circ$]  &  $\Omega$ [$^\circ$] &  $\sum \Delta V$ [m/s]  \\
      \hline  
      Titan unc. 500  &  $-332.6$ ($2,449.0$) & $21.09$ ($3.15$)  &  $30.13$ ($0.61$) & $89.95$ ($0.76$) &  $89.99$ ($0.05$) & $-$  ($-$)   \\
      \hline  
      Titan con. 500 &  $-128.1$ ($0.0$) & $25.02$ ($0.00$)  &  $30.00$ ($0.00$) & $90.00$ ($0.00$) &  $90.00$ ($0.00$) & $783.7$  ($106.7$)   \\
      \hline  
      Titan unc. 750 &  $-139.5$ ($5.5$) & $24.87$ ($1.00$)  &  $29.98$ ($0.60$) & $90.01$ ($0.10$) &  $90.00$ ($0.05$) & $-$  ($-$)   \\
      \hline  
      Titan con. 750 &  $-138.8$ ($0.0$) & $24.95$ ($0.02$)  &  $30.00$ ($0.00$) & $90.01$ ($0.01$) &  $90.00$ ($0.00$) & $212.7$  ($100.9$)   \\
      \hline  
      Enceladus unc. 10 &  $-2.65$ ($0.11$) & $100.60$ ($12.71$)  &  $29.78$ ($7.58$) & $89.98$ ($0.23$) &  $90.05$ ($0.24$) & $-$  ($-$)   \\
      \hline  
      Enceladus con. 10 &  $-2.64$ ($0.02$) & $99.99$ ($0.38$)  &  $30.00$ ($0.21$) & $90.00$ ($0.05$) &  $90.00$ ($0.05$) & $306.8$  ($219.9$)   \\
      \hline  
   \end{tabular}
\end{table}

\begin{figure}[!htb]
\centering
\subfloat[Trajectories]{\includegraphics[width=.5\textwidth]{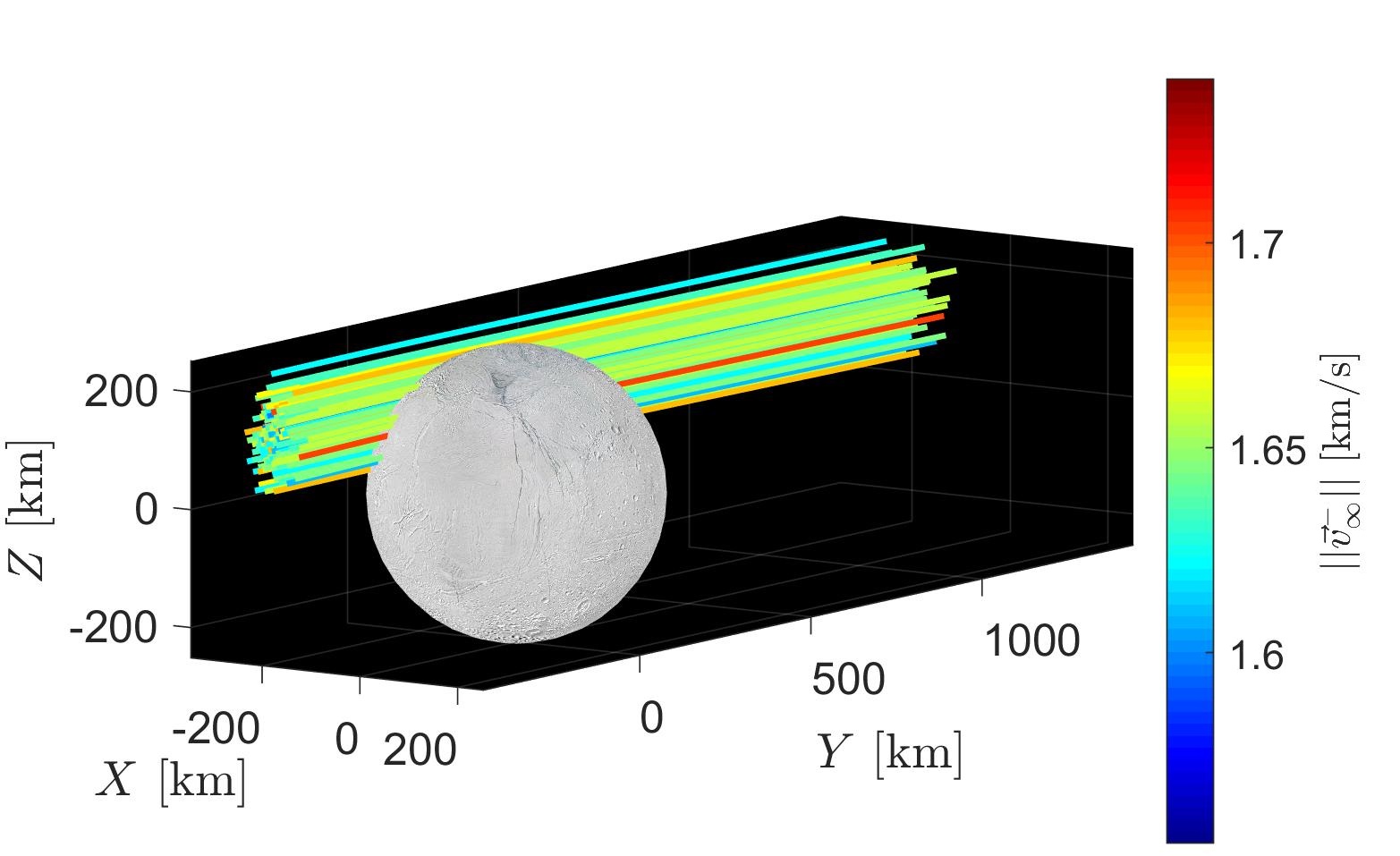}\label{fig:enceladus_un_3D}}\\
\subfloat[$a$ and $e$]{\includegraphics[width=.5\textwidth]{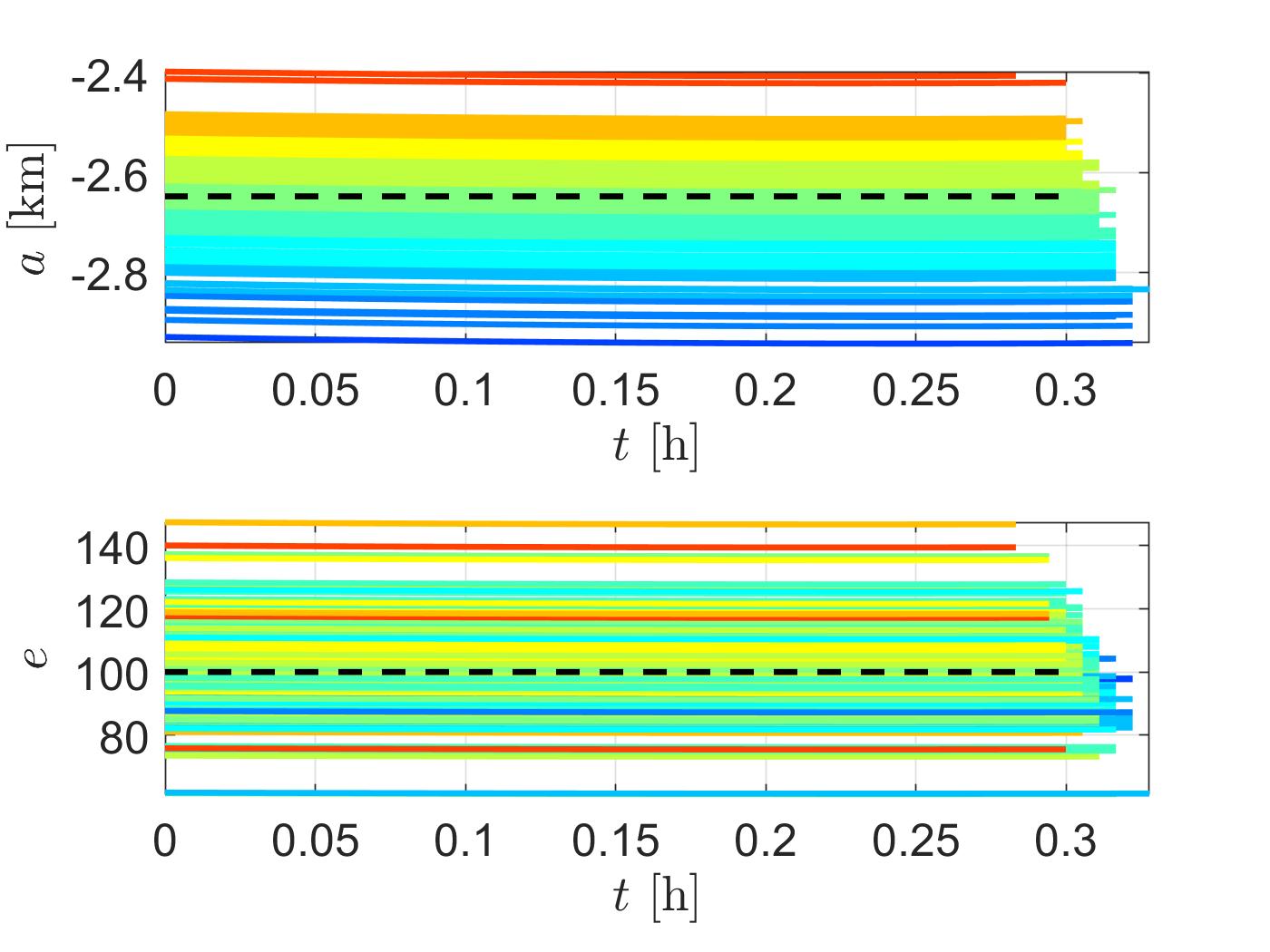}\label{fig:enceladus_un_ae}}  
\subfloat[$i$, $\omega$, and $\Omega$]{\includegraphics[width=.5\textwidth]{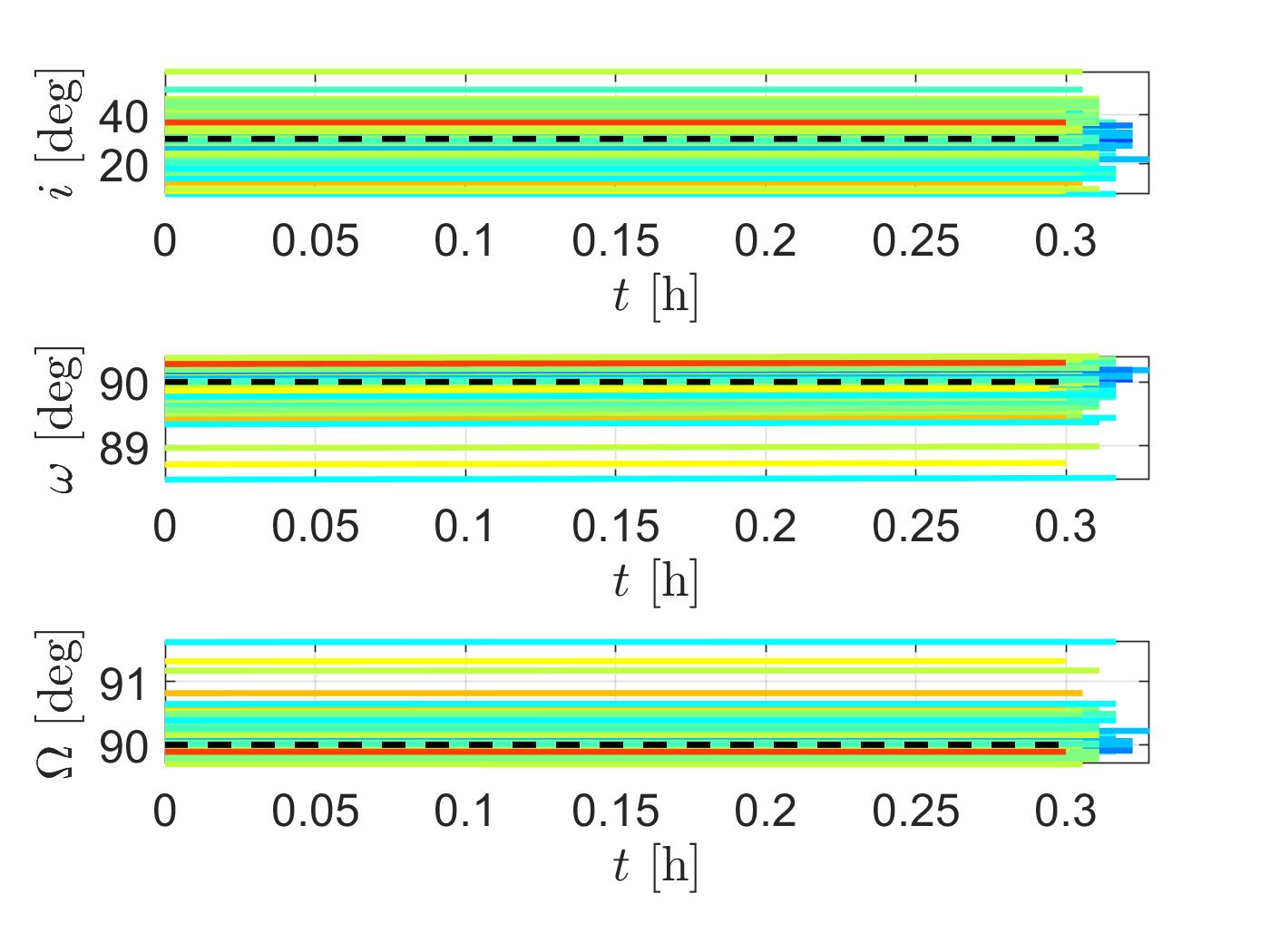}\label{fig:enceladus_un_iwO}}

\caption{Monte Carlo of the uncontrolled Enceladus' gravity-assist.}
\label{fig:enceladus_un}
\end{figure}

\begin{figure}[!htb]
\centering
\subfloat[Trajectories]{\includegraphics[width=.5\textwidth]{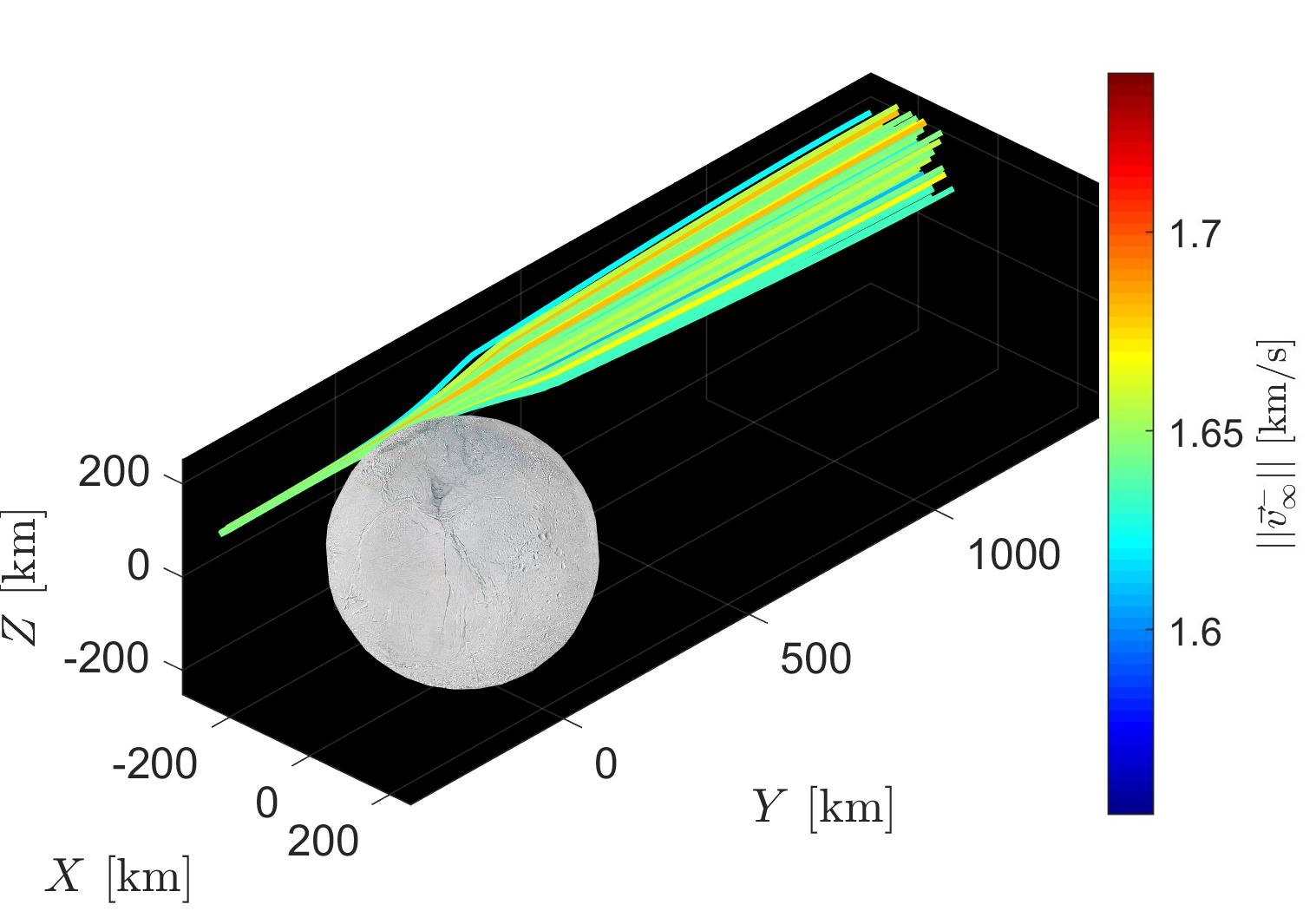}\label{fig:enceladus_con_3D}}
\subfloat[$\sum \Delta V$]{\includegraphics[width=.5\textwidth]{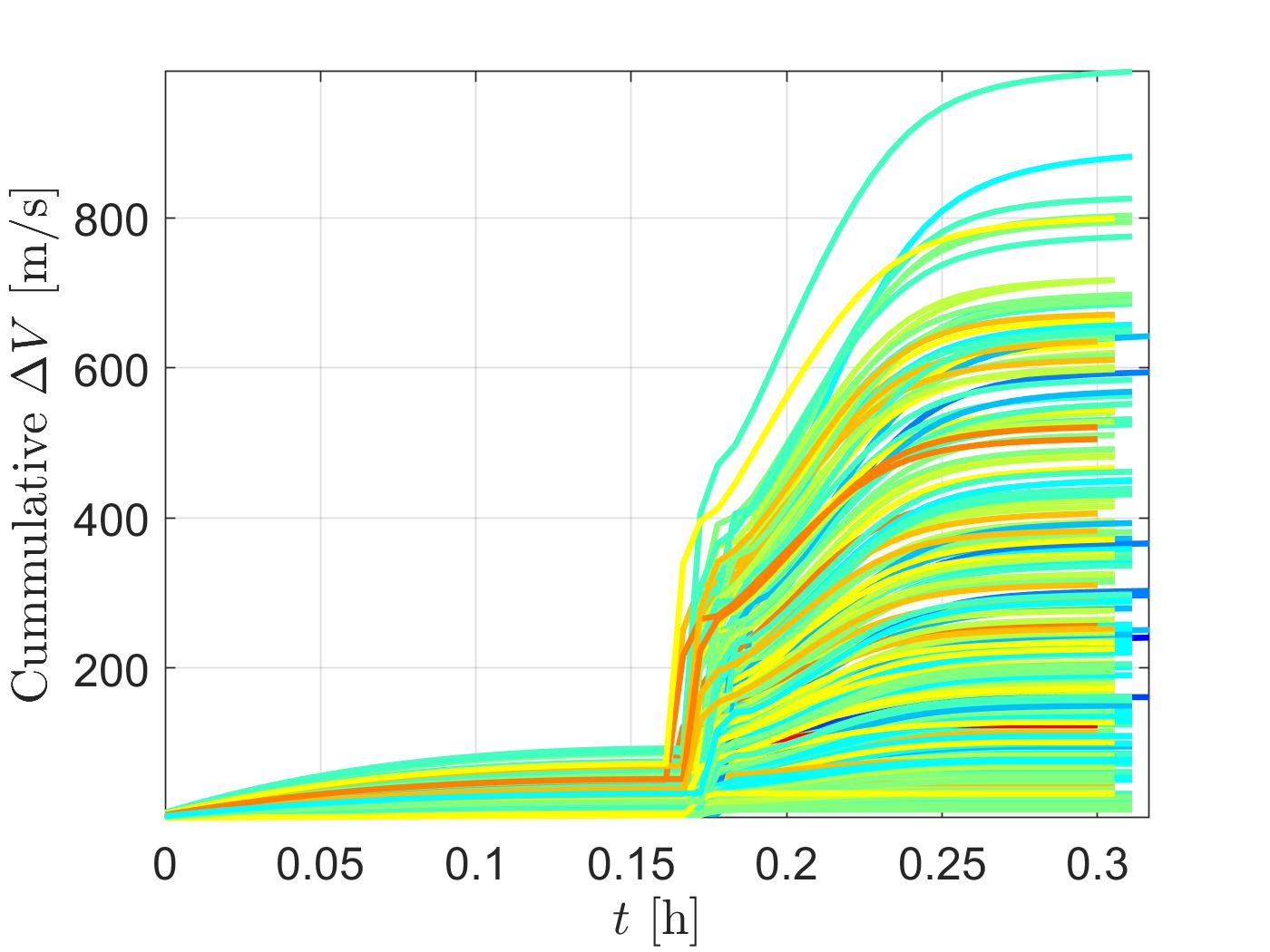}\label{fig:enceladus_con_dV}}  \\
\subfloat[$a$ and $e$]{\includegraphics[width=.5\textwidth]{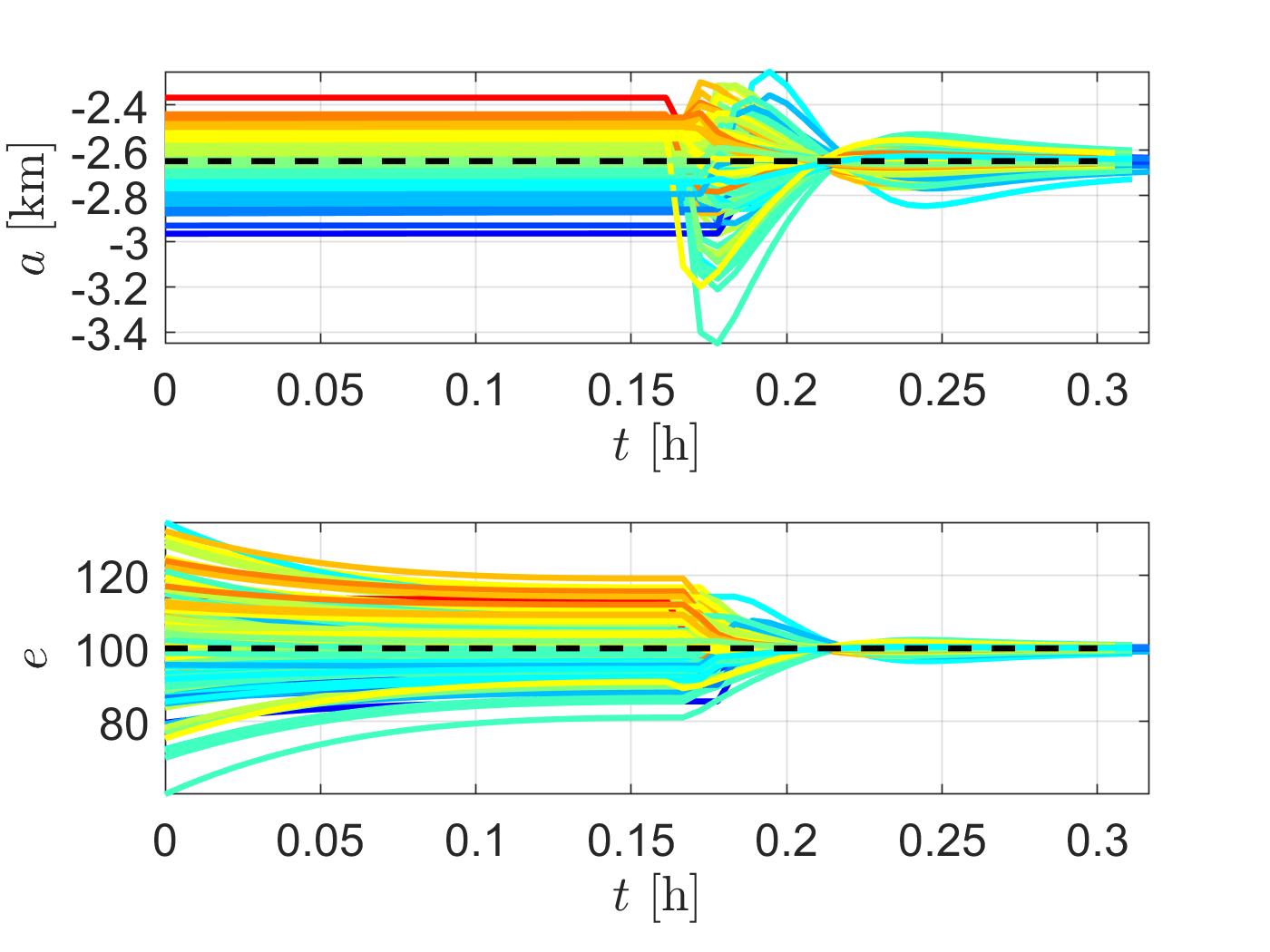}\label{fig:enceladus_con_ae}}  
\subfloat[$i$, $\omega$, and $\Omega$]{\includegraphics[width=.5\textwidth]{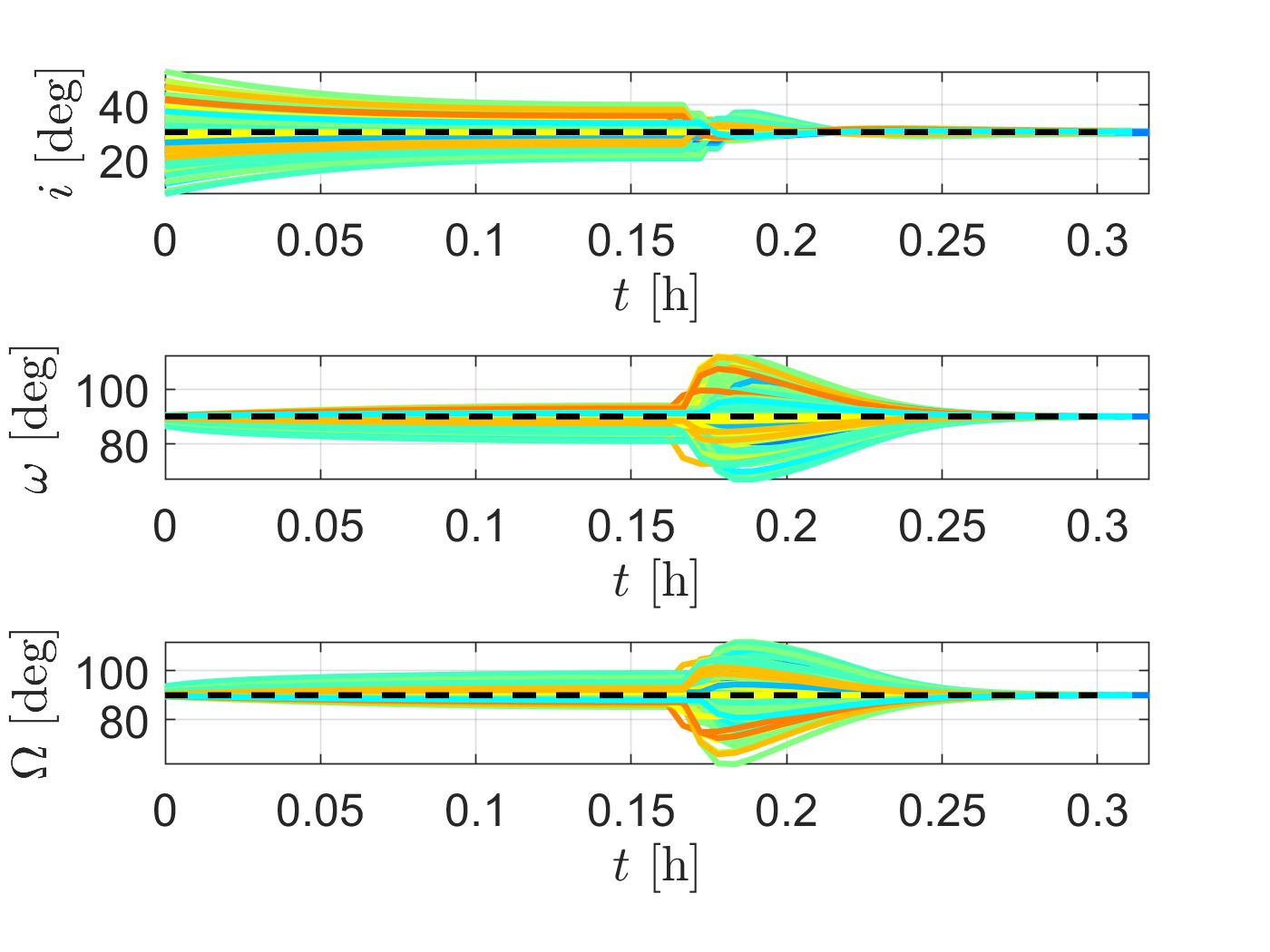}\label{fig:enceladus_con_iwO}}

\caption{Monte Carlo of the controlled Enceladus' gravity-assist.}
\label{fig:enceladus_con}
\end{figure}

Moving now to the small Saturnian moon Enceladus, for the same previous desired orbital elements, except for $e=100$ and an altitude about Enceladus of 10 km (below the closest Cassini approach of 25 km in October 2008). It is considered as disturbances the Saturn gravitational effects and the drag from Enceladus' plumes, with a mass density $\rho=$ 5.5$\times10^{-11}$ kg/m$^3$~\cite{lorenz2018enceladus}. The RKPFC parameters are adjusted to: $D_R=D_T=D_N=10.0$ m/s$^2$, $\lambda_R=\lambda_N=2$ and $\Phi$ is 30 times each corresponding element of $K$, with the switcher parameters: $\vec{\chi}^-= \begin{bmatrix} 10 & 1 & 0.1 & 0.1 & 0.1 \end{bmatrix}$ and $\vec{\chi}^+= \begin{bmatrix} 100 & 5.0 & 0.5 & 0.5 & 0.5 \end{bmatrix}$ (in meters or degrees). 

In Figure \ref{fig:enceladus_un} is shown the simulated trajectories with no control input, resulting in tens of collisions with Enceladus, Figure \ref{fig:enceladus_un_3D}. The largest dispersion in the orbital elements was observed for the eccentricity ($~20$) and the inclination ($~20^\circ$), Figures \ref{fig:enceladus_un_ae} and \ref{fig:enceladus_un_iwO}. In Figure \ref{fig:enceladus_con}, considering the proposed control, is shown the stabilized hyperbolic trajectory. Even in the drastic condition of having to be stabilized in a short time period, just a few Enceladus' radii from Enceladus surface (the SOI of Enceladus is $~1.93$ Enceladus' radii), the control proves its efficiency by rapidly stabilizing the orbit. The budget $\Delta V$ in this scenario, as can be checked in Table \ref{tab:label}, Enceladus con. 10, is of 306 m/s with a 1-$\sigma$ dispersion of 219.9 m/s. 

One might correctly wonder that the budget $\Delta V$ found for the previous simulations are indeed large, probably nullifying gravity-assist advantages. First, the proposed control scheme is advantageous for flybys in general, with no need of a gravity-assist. In the case of an asteroid or small moon flyby for scientific observation, the drastic circumstance of the presented Enceladus hyperbolic encounter indicates that the proposed control allow for a close approach with the target body, rapidly converging to the desired geometry. For instance, consider a hypothetical asteroid flyby. As the spacecraft finds the asteroid in its optical navigation cameras, a hyperbolic trajectory with the desired geometrical features (as a desired periapsis, and a convenient geometry for no need of a great change in the spacecraft's velocity) is instantaneously calculated. A good estimative of the distance and velocity relative to the target asteroid might only be available a few kilometers from the target, with a small-time for corrections. As the Enceladus example indicates, the proposed control might handle such a scenario. Secondly, as we previously pointed out, much of an autonomous gravity-assist depends on the guidance algorithm in calculating a trajectory as close to the real one as possible, avoiding such large insertion errors as the ones here considered. This is beyond the scope of this work, but the next section serves as a preliminary assessment, providing guidelines for future research.

\subsection{Jovian Tour}

Delay in the communications with a spacecraft in the Jovian system can span from 30 to 50 minutes, which compromises the ability to safely execute robust and efficient consecutive short transfer time flybys~\cite{quadrelli2015guidance}. So here we make a small analyses of the control applied to a Jovian system tour. We consider a tour by the Galilean moons, with the spacecraft departing from a position $X=-1.2\times 10^{6}$ km in the inertial frame, and having to arrive at the final moon of the tour with no other specified condition. A first optimal solution is obtained using the 0SOI-PC model with the MATLAB built-in genetic algorithm. The resulting optimal solution is used as an initial guess for a second optimization using the patched conics. The solution obtained in the PC optimization is considered as the guidance solution available for the spacecraft, and a simulation in the CRNBP is performed to assess the performance. 

\begin{figure}[!htb]
\centering
\subfloat[Approaches uncontrolled]{\includegraphics[width=.5\textwidth]{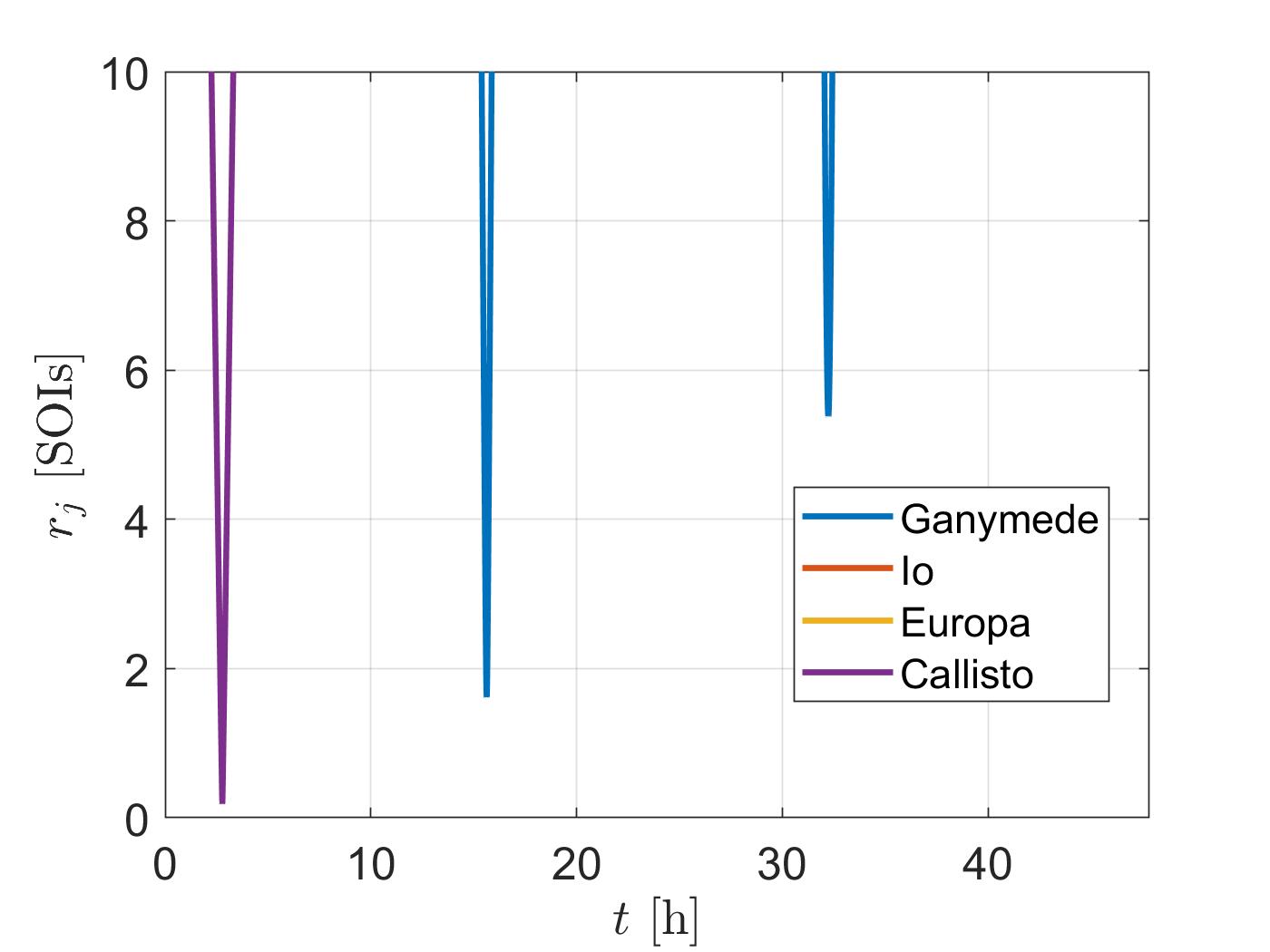}\label{fig:tour_CGEI_approaches_unc}} 
\subfloat[N-body uncontrolled]{\includegraphics[width=.5\textwidth]{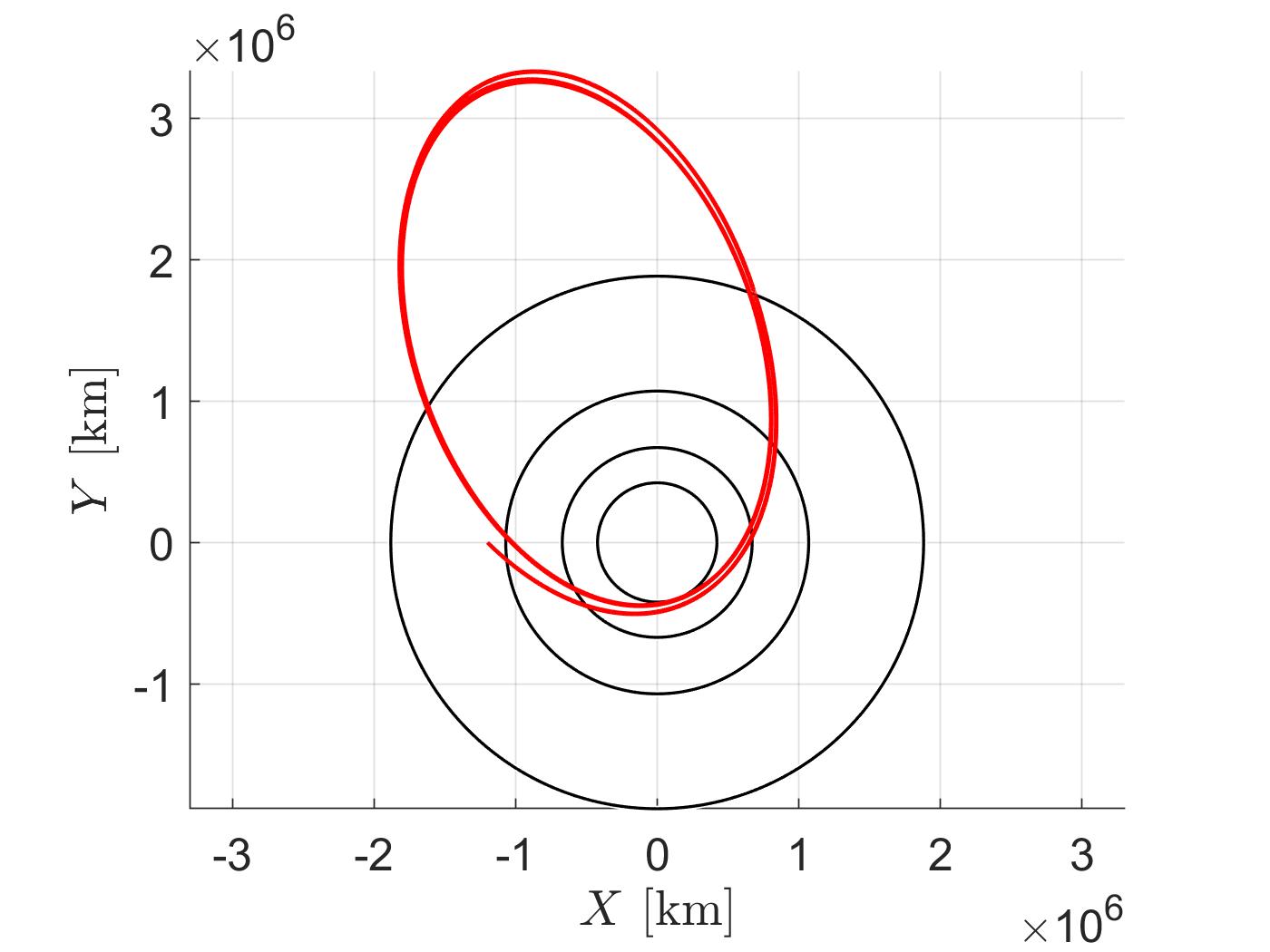}\label{fig:tour_CGEI_unc}}\\
\subfloat[Approaches controlled]{\includegraphics[width=.5\textwidth]{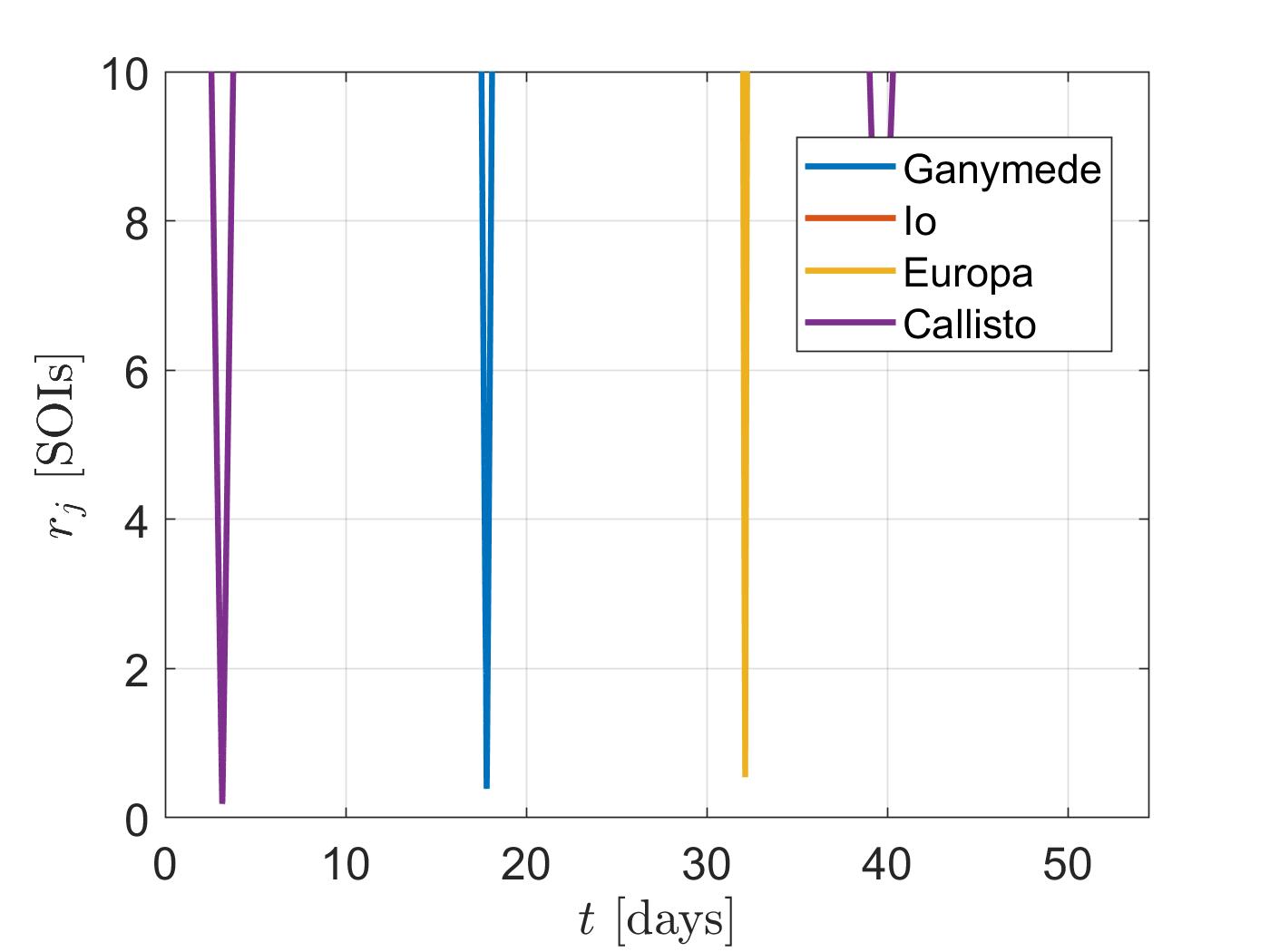}\label{fig:tour_CGEI_approaches_con}} 
\subfloat[N-body controlled]{\includegraphics[width=.5\textwidth]{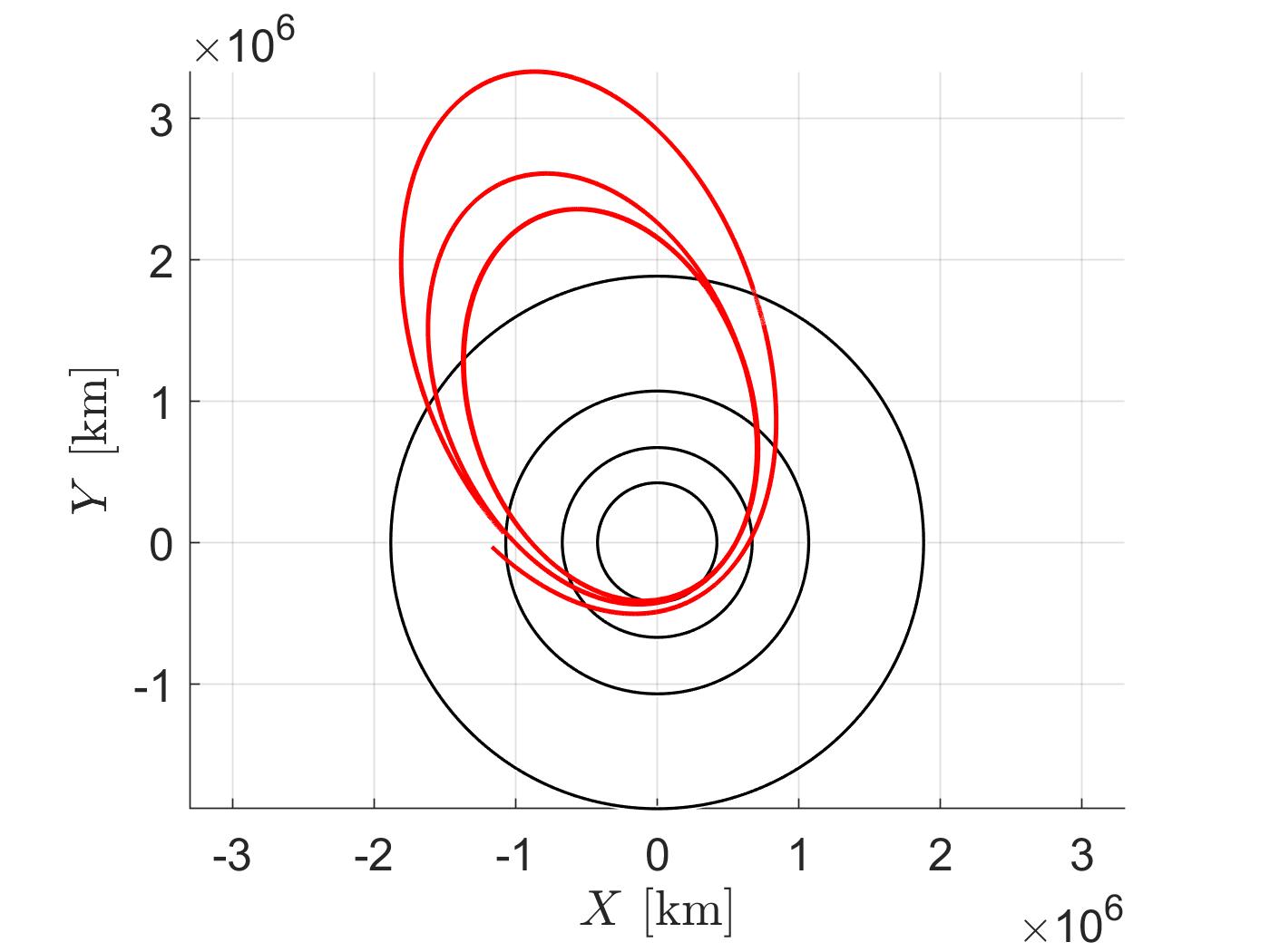}\label{fig:tour_CGEI_con}}\\
\subfloat[Patched conics]{\includegraphics[width=.5\textwidth]{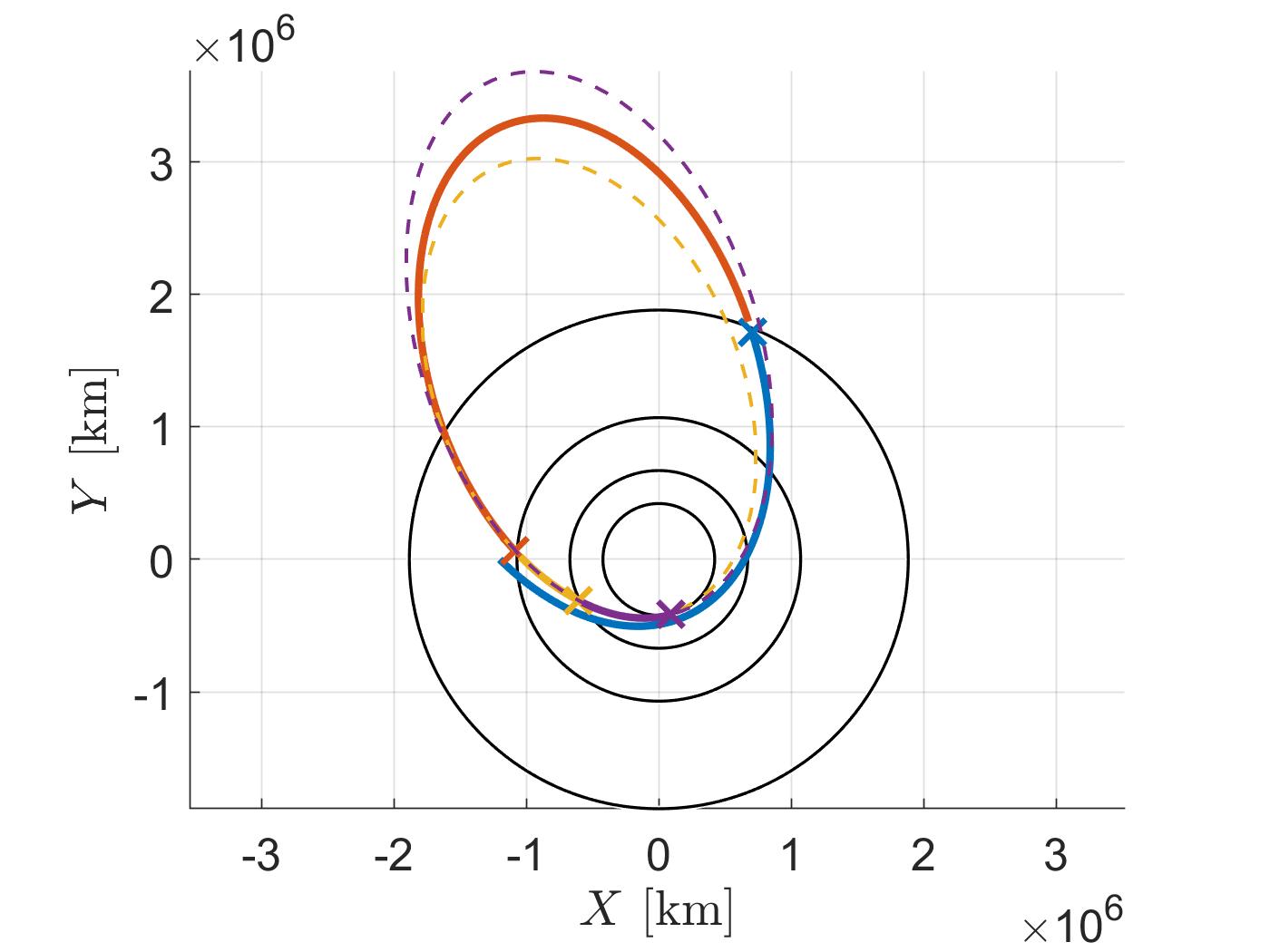}\label{fig:tour_CGEI_PC}}

\caption{Tour CGEI.}
\label{fig:tour_CGEI}
\end{figure}

Figure \ref{fig:tour_CGEI} presents the results obtained for a tour Callisto-Ganymede-Europa-Io obtained in the PC model as a free-fall, with periapsis encounters at the times: 3.1682, 17.8318, 34.0092 and   54.4843 days. The predicted trajectory with the PC is depicted in Figure \ref{fig:tour_CGEI_PC}, where dashed lines represent transfer orbits with more than an orbital period. The colors for the transfer ellipses are chosen, following the order of the transfers, as: blue, orange, yellow and purple. Figure \ref{fig:tour_CGEI_unc} shows the obtained optimal trajectory simulated in the CRNBP, with approaches to each of the bodies shown in Figure \ref{fig:tour_CGEI_approaches_unc}, normalized by the SOI radius of the respective moons. As one can note, the real trajectory largely diverge from the one obtained with the PC, with the encounter with Ganymede occurring outside the SOI, and no encounter with Europa at all.

Figures \ref{fig:tour_CGEI_approaches_con} and \ref{fig:tour_CGEI_con} present the case considering the control. The encounter with Ganymede in the desired conditions is guaranteed by the control. However, the encounter with Europa occurs before than expected, 32.1289 days, resulting in the divergence of the real trajectory to the nominal one thereafter. This type of behavior is quite common. The Jovian system is quite chaotic, we found no case in which a flyby control by itself can guarantee a tour with a guidance given by a patched conics solution. The $\Delta V$ budget is also quite prohibitive, in this tour example it amounts to 37.7 km/s.

These results indicate that for an autonomous spacecraft to operate in an outer planetary system, it is more likely that its trajectory is uploaded from time to time by a ground orbit determination team. The control would only compensate for small deviations, with the advantage of allowing consecutive shorter time transfers between flybys (which it is not the case for the nowadays missions). If the autonomy is also considered for the guidance, the spacecraft should be able to calculate the trajectory, with no supervision, probably running optimization routines, in a highly demanding and complex model.

\section{Conclusion}

We have proposed a robust path following control strategy for stabilizing flyby trajectories. Our analysis has demonstrated the effectiveness of this control approach even in challenging scenarios, such as approaching within Titan's atmosphere or performing a close-approach to Enceladus with limited response time. Furthermore, we have examined a Jovian tour trajectory calculated using a patched conics model, revealing the sensitivity and chaotic nature of outer planetary systems. Importantly, our results highlight that a flyby control strategy alone cannot guarantee the successful completion of a tour based on a patched conics model. Therefore, autonomous spacecraft operating under these conditions would require a high-fidelity guidance model, which could be periodically uploaded by a ground orbit determination team or implemented as an embedded model. These findings emphasize the additional challenges faced by autonomous operations in outer planetary systems and the need for precise guidance systems in such missions.

\section{Acknowledgment}
The authors wish to express their appreciation for the support provided by grants $\#$ 406841/2016-0 and 301338/2016-7 from the  National Council for Scientific and Technological Development (CNPq); grants $\#$ 2017/20794-2, 2015/19880-6 and 2016/24561-0 from S\~ao Paulo Research Foundation (FAPESP) and the financial support from the Coordination for the Improvement of Higher Education Personnel (CAPES).

%\appendix
%\section*{Appendix: Title here}
%Each appendix is its own section with its own section heading. The title of each appendix section is preceded by ``APPENDIX: '' as illustrated above, or ``APPENDIX A: '', ``APPENDIX B: '', \emph{etc}., when multiple appendixes are necessary. Appendices are optional and normally go after references; however, appendices may go ahead of the references section whenever the word processor forces superscripted endnotes to the very end of the document. The contents of each appendix must be called out at least once in the body of the manuscript.

%\subsection*{Miscellaneous Physical Dimensions}
%The page size shall be the American standard of 8.5 inches by 11 inches (216 mm x 279 mm). Margins are as follows: Top -- 0.75 inch (19 mm); Bottom -- 1.5 inches (38 mm); Left -- 1.25 inches (32 mm); Right -- 1.25 inch (32 mm). The title of the manuscript starts one inch (25.4 mm) below the top margin. Column width is 6 inches (152.5 mm) and column length is 8.75 inches (222.5 mm). The abstract is 4.5 inches (114 mm) in width, centered, justified, 10 point normal (serif) font.

\bibliographystyle{AAS_publication}   % Number the references.
\bibliography{references}   % Use references.bib to resolve the labels.

\end{document}